\documentclass[10pt,a4paper,twocolumn,english,aps]{revtex4-1}
\usepackage[T1]{fontenc}
\usepackage[latin9]{inputenc}
\usepackage{xcolor}
\usepackage{amsmath}
\usepackage{amssymb}
\usepackage{graphicx}
\PassOptionsToPackage{normalem}{ulem}
\usepackage{ulem}
\usepackage{nomencl}
\providecommand{\printnomenclature}{\printglossary}
\providecommand{\makenomenclature}{\makeglossary}
\makenomenclature

\makeatletter

\providecommand{\tabularnewline}{\\}
\providecolor{lyxadded}{rgb}{0,0,1}
\providecolor{lyxdeleted}{rgb}{1,0,0}
\DeclareRobustCommand{\lyxsout}[1]{\ifx\\#1\else\sout{#1}\fi}

\makeatother

\usepackage{babel}

\begin{document}
\title{Reversible stepwise condensation polymerization with cyclization:
strictly alternating co-polymerization and homopolymerization based
upon two orthogonal reactions}
\author{Michael Lang$^{1}$}
\email{lang@ipfdd.de}
\author{Kiran Suresh Kumar$^{1,2}$}
\affiliation{$^{1}$Institut Theorie der Polymere, Leibniz Institut für Polymerforschung
Dresden, Hohe Straße 6, 01069 Dresden, Germany}
\affiliation{$^{2}$Institut für Theoretische Physik, Technische Universität Dresden,
Zellescher Weg 17, 01069 Dresden, Germany}

\selectlanguage{english}%
\begin{abstract}
In a preceding work {[}M. Lang, K. Kumar, A simple and general approach
for reversible condensation polymerization with cyclization, \emph{Macromolecules}
\textbf{54} (2021), in press. ma-2021-00718y{]}, we have introduced
a simple recursive scheme that allows to treat stepwise linear reversible
polymerizations of any kind with cyclization. This approach is used
to discuss the polymerization of linear Gaussian strands (LGS) with
two different reactive groups $A$ and $B$ on either chain end that
participate in two orthogonal reactions and the strictly alternating
copolymerization of LGS that carry $A$ reactive groups with LGS equipped
with type $B$ reactive groups. The former of these cases has not
been discussed theoretically in literature, the latter only regarding
some special cases. We provide either analytical expressions or exact
numerical solutions for the general cases with and without cyclization.
Weight distributions, averages, polydispersity, and the weight fractions
of cyclic and linear species are computed. All numerical solutions
were tested by Monte-Carlo simulations.
\end{abstract}
\maketitle

\section{Introduction}

Polymers with dynamic bonds are interesting materials for many applications
as the material properties can be triggered by external stimuli \citep{McBride2019}.
New functionalities like the ability to self-heal \citep{Campanella2018}
or easy routes for recycling \citep{Hodge2014,Bapat2020} can be implemented,
while simultaneously, the material properties can be optimized regarding
the particular demands of highly specialized applications \citep{Zhang2018}.

Linear step growth polymerization is one of the classical routes to
prepare supramolecular polymers. One crucial point is there the formation
of cyclic molecules along with linear chains \citep{Flory1953}, which
complicates analysis and prediction of the material properties since
cyclic molecules exhibit different dynamics \citep{Kapnistos2008,Michieletto2016}
and conformations \citep{Grosberg1996,Lang2012} as their linear counterparts.
In particular, mixtures of both architectures \citep{Zhou2019} or
samples composed of molecules with largely different weights \citep{Lang2015a,Lang2013b}
may develop a quite complex behavior that can be sensitive regarding
traces of molecules with a different architecture \citep{Kapnistos2008}.
Therefore, one key for understanding the material properties is an
accurate model for composition and weight distributions of the linear
and cyclic molecules. It is the aim of the present work to provide
such a model for two special cases of a linear step growth polymerization.

\begin{figure}
\includegraphics[width=1\columnwidth]{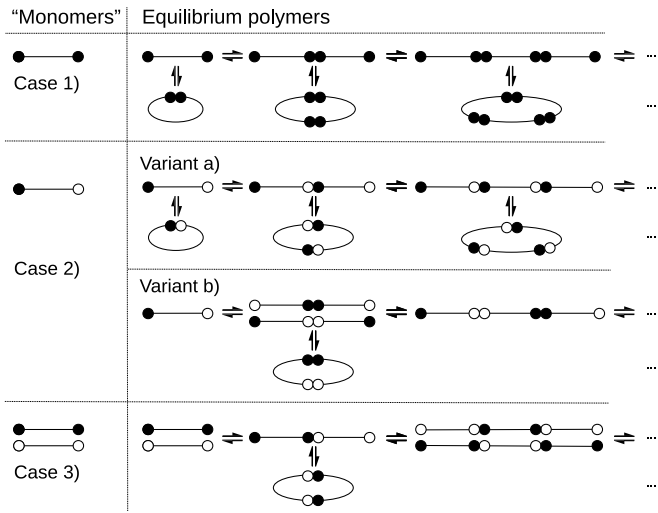}

\caption{\label{fig:The-three-classical}The three classical cases of linear
polymerization discussed in Ref. \citep{Jacobson1950} that involve
only two different reactive groups and no more than two different
macromonomers. Reactive groups of a different type are displayed by
different beads. The different macromonomers on the left assemble
into linear and cyclic molecules where the simplest ones are shown
on the right.}
\end{figure}

In our preceding paper \citep{Lang2021a}, we have developed a simple
framework to treat such kind of polymerizations and tested it for
two classes of step growth polymerization (case 1 and case 2a, see
Figure \ref{fig:The-three-classical} for a sketch of these reactions).
In the present work, we apply this approach to the remaining two cases
of a reversible linear step growth polymerization shown in Figure
\ref{fig:The-three-classical}. Historically, \citep{Jacobson1950},
only three different cases were distinguished, since by the time when
Jacobson and Stockmayer (JS) published their seminal work, systems
with two orthogonal reactions were unknown. In these orthogonal systems,
monomers have two different chain ends of type $A$ and $B$ respectively
that react only with like reactive groups. Since this is the complementary
case to the original case 2, we call this case 2b. The second type
of reaction that we treat in the present work is called case 3 and
refers to a strictly alternating sequence of $A$ terminated macromonomers
with $B$ terminated macromonomers. Note that we call these macromonomers
``strands'', if we talk about single precursor units. The term ``molecule''
is used for assemblies of $k=1,2,3,...$ strands. If architecture
of the molecules matters, we distinguish (linear) chains from ``cyclic
molecules'', that are called ``rings'' or ``loops'' for the sake
of brevity.

The classical example for case 3 is the reaction of adipic acid with
decamethylene glycole \citep{Flory1936,Jacobson1950}, more recent
examples include the association of diaminotriazine with thymine stickers
\citep{Bras2013} and most linear metallo-supramolecular chain extended
polymers like Ref \citep{Mansfeld2013} form alternating sequences
of two units and thus, fall into this category. Several examples for
the orthogonal reactions of case 2b can be found in Refs. \citep{Hofmeier2005,Groeger2011,Li2012}.
Reactions of this latter type have attracted significant attention
in recent years, since two independent mechanisms can be addressed
by an external stimulus. These developments have also found application
in the construction of multi stimuli-responsive networks \citep{Qian2016,Sataux2018}
or hyperbranched polymers \citep{Gu2015}.

Once supramolecular bonds establish, one is confronted with the problem
of characterizing the supramolecular polymers. This is not a simple
task at the best of times, as the molecules may re-assemble on the
time scale of the experiment \citep{Moratti2005}. Similar to covalently
linked polymers, a characterization of the supramolecular polymers
requires some insight into the average molecular weights, as these
are probed by different experimental techniques. Often, not only the
average molecular weight, but also the distribution and its width
are essential for properties of the polymer material \citep{Gentekos2019}.
Therefore, a precise prediction of these quantities is of a large
interest to understand the behavior of supramolecular polymers.

An irreversible alternating co-polymerization without loop formation
was partially treated in Flory's original work \citep{Flory1936}
on condensation polymerization omitting a computation of the weight
averages and polydispersity of both, the differently terminated chains
and the full sample. A later attempt to provide the missing averages
\citep{Mizerovskii2013} was not successful, as discussed in the Appendix.
Furthermore, weight distributions and averages for case 2b without
loop formation were not discussed in literature to the best of our
knowledge. We close this gap by deriving the corresponding distributions
and averages for the loop free limit in the Appendix.

With consideration of loop formation, case 3 was discussed only for
stoichiometrically balanced systems or completely reacted minority
species in the original work of JS \citep{Jacobson1950}. One may
recall here also the limitations of the the JS approach, that provides
no quantitative prediction for conversion, etc., see Ref. \citep{Lang2021a}
for a more detailed discussion. Random co-polymerization in the presence
of cyclization has been discussed by Szymanski \citep{Szymanski1989,Szymanski1992}
without covering the case of an alternating co-polymerization. Vermonden
et al. \citep{Vermonden2003} applied the JS model to case 3, addressing
ring-chain equilibria in strictly alternating systems of water soluble
coordination polymers. Here, the second ligand complex with a metal
ion yields a different binding energy, which leads to asymmetric results
that can be modeled as a first shell substitution effect. Note that
the treatment by Vermonden et al. \citep{Vermonden2003} is based
upon sample average probabilities. However, cyclic molecules with
an alternating sequence of building block have always a balanced stoichiometry,
see section \emph{Alternating co-polymerization (case 3)}. Deviations
from stoichiometry are balanced within the linear species alone. Therefore,
the treatment in Ref. \citep{Vermonden2003} is only approximate and
becomes increasingly inaccurate for an increasing weight fraction
of rings or stoichiometric imbalance. To the best of our knowledge,
there is no accurate and self-consistent treatment of case 2b and
3 available in literature no matter whether cyclization is included
or not. It is the aim of the present paper, to provide this treatment
in its simplest form focusing on linear Gaussian strands (LGS) as
basic building blocks and using Flory's simplifying assumption of
equal reactivity of reactive groups of the same kind and independence
of reactions.

\nomenclature{LGS}{linear Gaussian Strands}%

\nomenclature{JS}{Jacobson and Stockmayer}%

In the following sections, we extend the approach of Ref. \citep{Lang2021a}
to case 2b and case 3, whereby we start with the latter to simplify
the discussion. We use the numerical scheme that is explained in the
appendix of Ref. \citep{Lang2021a} to obtain exact numerical solutions
of the set of balance equations. Note that an example for this scheme
is given in the SI of Ref. \citep{Lang2021a}. Also, the second section
of Ref. \citep{Lang2021a} is a useful introduction for our approach,
since it contains the basic expressions for intra- and intermolecular
reactions, the law of mass action, and the balance equations that
are applied below. All key findings related to weight fractions of
rings or weight distributions of rings and linear chains are tested
by Monte Carlo simulations. These are also described in the Appendix
of Ref. \citep{Lang2021a}.

\section{\label{subsec:Alternating-linear-co-polymeriza}Alternating co-polymerization
(case 3)}

Let us consider the case of an alternating polymerization where $2$-functional
strands of type $A$ react exclusively with $2$-functional strands
of type $B$. Let 
\begin{equation}
r=\frac{c_{\text{A}}}{c_{\text{B}}}\label{eq:r}
\end{equation}
denote the stoichiometric ratio of the concentrations of reactive
groups, $c_{\text{A}}$ and $c_{\text{B}}$, of strands of type $A$
and $B$ respectively. The total concentration of reactive groups
is here 
\begin{equation}
c_{\text{t}}=c_{\text{A}}+c_{\text{B}}=c_{\text{B}}\left(r+1\right).\label{eq:ct}
\end{equation}
Once $r\ne1$ for irreversible systems, one typically assumes that
the minority species is converted completely, while non-reacted groups
are located exclusively on the majority species \citep{Suckow2019}.
For reversible systems, such an assumption is not feasible as unbound
groups are continuously created by bond breaking. Without loss of
generality, let us choose $A$ as the minority component, which restricts
our discussion to $r<1$. We further simplify the discussion by assuming
that the strands $A$ and $B$ are identical except for the end groups,
so that both strands occupy roughly the same volume.

\nomenclature{$c_{\text{t}}$}{total concentration of reactive groups}%

\nomenclature{$c_{\text{X}}$}{concentration of reactive groups of type $X=A,B$}%

\nomenclature{$A, B$}{type of reactive group}%

\nomenclature{$r$}{stoichiometric ratio between $A$ and $B$ groups}%

To proceed, we require the number fraction distribution of linear
species with an even number of strands, since only these can form
loops in an alternating co-polymerization. Virtually all treatments
of linear or non-linear co-polymerization do not distinguish between
chains with an even or odd number of strands. Instead, they focus
mainly on average molecular weights, as these are easier to derive,
see e.g. \citep{Stockmayer1944,Flory1946,Flory1953,Macosko1976}.
The only exceptions we could find are Refs. \citep{Flory1936,Mizerovskii2013}.
Even there, not all required distributions and averages are available.
Quite surprisingly, none of these works provides a correct set of
equations for the number and weight distributions as can be shown
by checking for normalization. Therefore, we added section \emph{Case
3 without rings} to the Appendix, where we present a complete derivation
of all required distributions and averages.

To model cyclization, we must distinguish the chains regarding their
ends. We call chains ``$A$-terminated'', if two $A$ strands are
on their ends, ``$B$ terminated'' chains have two ends of type
$B$, while ``$m$-terminated'' chains have end groups of both types.
We use an index $A$, $B$, or $m$ to indicate that a distribution
or averages refers to one of these particular classes of chains. Weight
fractions are denoted as $\text{w}$, while number fractions are denoted
by $n$. Thus, $\text{w}_{\text{A}}$ is the weight fraction of $A$-terminated
chains, while $n_{\text{m}}$ is the number fraction of mixed terminated
chains, as an example. Below, we use also a second kind of weight
fractions $\text{w}_{\text{j}}$ where $j=0,1,2$ counts the number
of bound reactive groups (``closed stickers'' \citep{Stukhalin2013})
of the strands. For distinction, the total weight fraction of loops
is denoted by $\omega$ while weight fractions of loops made of $k$
strands is written as $\omega_{\text{k}}$. Finally, $\omega_{\text{A}}$
is the weight fraction of $A$ strands that are part of loops.

\nomenclature{w$_{\text{X}}$}{weight fraction of $X=A,B,m$ terminated chains}%

\nomenclature{$m$}{index for mixed terminated chains}%

\nomenclature{$n_{\text{X}}$}{number fraction of $X=A,B,m$ terminated chains}%

\nomenclature{w$_{\text{j}}$}{weight fraction of strands with $j$ stickers closed}%

\nomenclature{$\omega$}{weight fraction of loops}%

\nomenclature{$\omega_{\text{k}}$}{weight fraction of loops made of $k$ strands}%

\nomenclature{$\omega_{\text{A}}$}{weight fraction of $A$ strands in loops}%

\nomenclature{$k$}{number of strands per molecule}%

\nomenclature{$p_{\text{lin}}$}{conversion of chain fraction}%

\nomenclature{$p$}{conversion of reactive groups}%

\nomenclature{$r_{\text{lin}}$}{stoichiometric ration in the chain fraction}%

In case of loop formation, the loops are always at 100\% conversion,
contain an even number of strands, and must be stoichiometrically
balanced due to the alternating scheme of case 3. Similar to equation
(16) of Ref. \citep{Lang2021a}, loop formation reduces the conversion,
$p$, inside the linear chain species to
\begin{equation}
p_{\text{lin}}=\frac{p-\omega_{\text{A}}}{1-\omega_{\text{A}}}.\label{eq:p-lin again}
\end{equation}
Furthermore, the stoichiometric balance of the loops, shifts the stoichiometric
ratio of the linear fraction to 
\begin{equation}
r_{\text{lin}}=\frac{r\left(1-\omega_{\text{A}}\right)}{1-r\omega_{\text{A}}}.\label{eq:r_lin}
\end{equation}
Note that $\omega_{\text{A}}$ enters here in both equations above
instead of $\omega$ that was used in Ref. \citep{Lang2021a}, since
the total weight fraction of loops is limited by the weight fraction
of the minority species $A$.

Both $p_{\text{lin}}$ and $r_{\text{lin}}$ describe the properties
of the linear chain fraction in the presence of loops, and replace
$p$ and $r$ in all equations that are taken from section \emph{Case
3 without rings} of the Appendix. To clarify this point in our notation,
we add to all quantities taken from the Appendix the additional suffix
``lin''. Furthermore, we have computed all weight fractions $\text{w}_{\text{X}}$
with $X=A,B,m,...$ etc. in section \emph{Case 3 without rings} in
the absence of loops. Normalization of these quantities with respect
to the full sample is obtained by multiplication with $1-\omega$.

\nomenclature{$w_{\text{X,0,lin}}$}{weight fraction of smallest $X=A,B,m$ terminated molecule}%

We proceed as in Ref. \citep{Lang2021a} by proposing balance equations
for $\text{w}_{\text{j}}$. Since any reaction of an $A$ group involves
a reaction of a $B$ group, it is sufficient to write down the balance
equations only in terms of the $A$ groups skipping an additional
suffix $A$ for all $\text{w}_{\text{j}}$. The weight fraction of
non-reacted $A$ strands, $\text{w}_{0}$, must be part of the weight
fraction of linear chains, $1-\omega$, and is given by 
\begin{equation}
\text{w}_{0}=\text{w}_{\text{\text{A,0}}}\left(1-\omega\right)=\frac{r_{\text{lin}}\left(1-p_{\text{lin}}\right)^{2}}{1+r_{\text{lin}}}\left(1-\omega\right),\label{eq:w0}
\end{equation}
see equation (\ref{eq:omegaa}) of the Appendix for $\text{w}_{\text{A,0}}$.

For the balance equation of strands $\text{w}_{1}$ with strands $\text{w}_{0}$,
we have to consider that the concentration of reaction partners of
type $B$ is $\left(1-rp\right)c_{\text{B}}$. Furthermore, there
are two chain ends of $\text{w}_{0}$ that can react, while the law
of mass action, equation (\ref{eq:K-1}), does not contribute another
factor of two in contrast to case 1 or 2b. Altogether, we obtain 
\begin{equation}
\text{w}_{\text{1}}=2Kc_{\text{B}}\left(1-rp\right)\text{w}_{0}=\frac{2Kc_{\text{t}}}{r+1}\left(1-rp\right)\text{w}_{0}.\label{eq:w1-1}
\end{equation}
Regarding the balance between $\text{w}_{2}$ and $\text{w}_{1}$,
we consider first only those forward reactions that do not lead to
cyclization and only backwards reactions where no cyclic molecule
transforms into a linear chain. Therefore, we put only the weight
fraction of $A$ strands that are not in cycles on the left hand side,
$\text{w}_{2}-\omega_{\text{A}}$, together with a symmetry factor
of two that reflects that each strand $\text{w}_{2}$ contributes
to two bonds that can break, while only one reactive group of $\text{w}_{1}$
can form bonds. In analogy to equation (15) of Ref. \citep{Lang2021a},
we obtain
\begin{equation}
2\left(\text{w}_{2}-\omega_{\text{A}}\right)=Kc_{\text{B}}\left(1-rp\right)\text{w}_{\text{1}}=\frac{Kc_{\text{t}}}{\left(r+1\right)}\left(1-rp\right)\text{w}_{\text{1}}.\label{eq:w2}
\end{equation}

Loop formation of the smallest ring does not couple to $w_{0}$ as
in case 1 polymerization, instead, it couples to the concentration
of dimers. These establish a weight fraction of
\begin{equation}
\text{w}_{\text{m,0,lin}}=\frac{4r_{\text{lin}}p_{\text{lin}}\left(1-p_{\text{lin}}\right)\left(1-r_{\text{lin}}p_{\text{lin}}\right)}{1+r_{\text{lin}}}\left(1-\omega\right)\label{eq:omolin}
\end{equation}
among all molecules, see equation (\ref{eq:omegam}).

\nomenclature{$z$}{number of addition pairs of strands beyond smallest chain with same end-groups}%

\nomenclature{$c_{\text{i}}$}{concentration of intramolecular reactive groups}%

\nomenclature{$c_{\text{e}}$}{concentration of external reactive groups}%

The concentration of the second dimer end next to the first is $2^{-3/2}c_{\text{i}}$.
Here, $c_{\text{i}}$ is the concentration of the second end around
the first of a single LGS, see equation (10) for $f=2$ of Ref. \citep{Lang2021a}.
Since there are two bonds per cyclic dimer that can break, we obtain
for the weight fraction of the smallest loop the balance equation
\begin{equation}
2\omega_{1}=2^{-3/2}c_{\text{i}}K\text{w}_{\text{m,0,lin}}.\label{eq:2w1}
\end{equation}

In total, this leads to an extra coefficient of $2^{-1/2}$ for $\omega_{1}$
as compared to case 1. Longer chains that can form loops contain $z$
additional pairs of $A$ and $B$ strands as compared to the dimer,
and they exist with a reduced probability $\left(r_{\text{lin}}p_{\text{lin}}^{2}\right)^{z}$,
see section \emph{Case 3 without rings}. As for case 1 discussed in
Ref. \citep{Lang2021a}, this leads to a total weight fraction of
$A$-mers in the loops that is a function of the smallest loop, $\omega_{1}$,
\begin{equation}
\omega_{\text{A}}=\frac{\omega_{1}}{2}\sum_{z=1}^{\infty}z^{-3/2}\left(r_{\text{lin}}p_{\text{lin}}^{2}\right)^{z-1}.\label{omega}
\end{equation}
The extra coefficient of $1/2$ in this equation reflects the fact
that only one half of all strands in the loop is of type $A$. The
total weight fraction of loops among all $A$ and $B$ strands is
therefore
\begin{equation}
\omega=\frac{2\omega_{A}}{1+r}.\label{eq:omega-1}
\end{equation}

In the limit of $c_{\text{t}}\gg c_{\text{i}}$ and $K\rightarrow\infty$
where $r_{\text{lin}}p_{\text{lin}}^{2}\rightarrow1$, we obtain for
$r=1$ a shift of the critical concentration, $c_{\text{crit}}$,
by a factor of $2^{-1/2}$ towards smaller concentrations as compared
to case 1, 
\begin{equation}
c_{\text{crit}}=2^{-1/2}\sum_{z=1}^{\infty}z^{-3/2}c_{\text{i}}.\label{eq:c_crit}
\end{equation}
This shift results from a factor of $2^{-3/2}$ due to end-contacts
of dimers instead of monomers and a factor of 2 for $r=1$ regarding
the concentration of possible reaction partners. The number density
of loops per strand is
\begin{equation}
n_{\text{L}}=\frac{\omega_{1}}{2\left(1+r\right)}\sum_{z=1}^{\infty}z^{-5/2}\left(r_{\text{lin}}p_{\text{lin}}^{2}\right)^{z-1},\label{eq:number density}
\end{equation}
which provides the number average degree of polymerization (DP) of
the loops through
\begin{equation}
N_{\text{n,loops}}=\frac{\omega}{n_{\text{L}}}.\label{eq:Nnloops}
\end{equation}

\nomenclature{DP}{degree of polymerization}%

\nomenclature{$n_{\text{L}}$}{number density of loops per strand}%

\nomenclature{$N_{\text{n}}$}{number average DP}%

\nomenclature{$N_{\text{n,lin}}$}{number average DP of chains}%

\nomenclature{$N_{\text{n,loops}}$}{number average DP of loops}%

\nomenclature{$N_{\text{w}}$}{weight average DP}%

As discussed in Ref. \citep{Lang2021a}, the above equations together
with the normalization of $\text{w}_{\text{j}}$ and the definition
of $p$ given in the Appendix of Ref. \citep{Lang2021a} allow to
solve the set of balance equations numerically. With the solution
of these equations, $\omega_{\text{A}}$ and $p$ become available,
which is the basis for computing the missing distributions and averages
of linear chains as described below for arbitrary $p$ and $r$. This
provides a significant advancement as compared to previous work. JS
\citep{Jacobson1950} discuss case 3 polymerization only in the limits
of a) $r=1$ while $p\ne1$ and b) $p=1$ while $r\ne1$. Vermonden
et al. \citep{Vermonden2003} apply the JS model to water soluble
coordination polymers. Similar to JS, these authors do not consider
that ring formation reduces the conversion in the linear chain fraction
and that ring formation increases the stoichiometric imbalance of
the linear chains, see e.g. equation (7) of Ref. \citep{Vermonden2003},
where only sample average quantities ($p$ and $q$ in their notation)
enter. This neglect allows to solve the set of equations without a
recursion, but affects the accuracy of their model once a significant
amount of rings is formed or a significant stoichiometric imbalance
is obtained.

\begin{figure}
\includegraphics[angle=270,width=1\columnwidth]{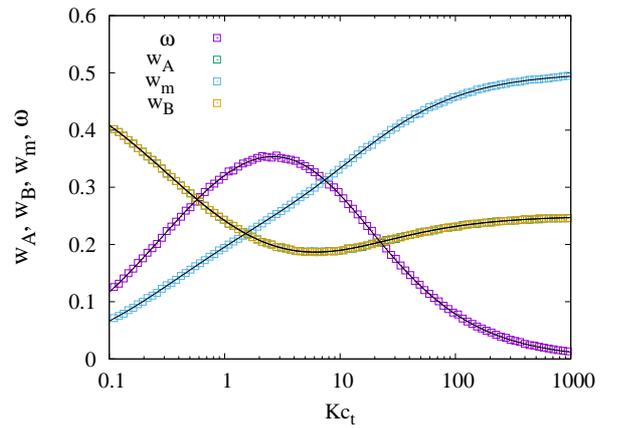}

\caption{\label{fig:Weight-fractions-of}Weight fractions of $A$, $B$, and
$m$-terminated chains, and loops for $c_{\text{i}}=10^{-2}$ as a
function of $Kc_{\text{t}}$ for stoichiometric ratio $r=1.0$ and
$K=10^{3}$. Symbols refer to simulation data, lines are numerical
solution of the balance equations for case 3.}
\end{figure}

For the presentation of the most relevant dependencies on reaction
constant(s) and concentrations, we have chosen a similar parameter
range as in our preceding work \citep{Lang2021a}, see the detailed
discussion there. In experiments, the reaction constant can be adjusted
with the temperature of the sample, see equation (24) of Ref. \citep{Lang2021a},
or by choosing a different chemistry for the reactive groups. However,
care needs to be taken here as many physical parameters (interactions
between the molecules, viscosity, ...) are a function of temperature.
The temperature dependence of these parameters might interfere largely
with the desired modification of the reaction constant.

The weight fractions of $A$, $B$, $m$-terminated chains and the
weight fraction of loops are shown in Figure \ref{fig:Weight-fractions-of}.
In marked contrast to case 1 polymerization, see Ref. \citep{Lang2021a},
there is a maximum of loop formation that precedes the maximum ($r<1$)
or the approach of saturation ($r=1)$ of the weight fraction of the
mixed terminated chains
\begin{equation}
\text{w}_{\text{m}}=\frac{4r_{\text{lin}}p_{\text{lin}}(1-p_{\text{lin}})(1-r_{\text{lin}}p_{\text{lin}})}{\left(1+r_{\text{lin}}\right)\left(1-r_{\text{lin}}p_{\text{lin}}^{2}\right)^{2}}\left(1-\omega\right)\label{eq:omegam renormalized}
\end{equation}
for increasing $Kc_{\text{t}}$, since loops are derived predominantly
from the shortest chains of $\text{w}_{\text{m}}$. These shortest
chains are dimers containing one intermolecular bond that disassembles
in the limit of very low concentration. On the other hand, the probability
for loop formation decreases in the limit of high concentrations.
In between these limits, there is an optimum concentration for loop
formation regarding the weight fraction of loops (but not regarding
the total weight of loops in the sample, see section \emph{Discussion}).

In Figure \ref{fig:Weight-fractions-of}, the data for $\text{w}_{\text{A}}$
and $\text{w}_{\text{B}}$ coincide due to symmetry. This Figure shows
also that the limit of low concentrations, $c_{\text{t}}\rightarrow0$,
refers to the limit of $p\rightarrow0$, where isolated linear macromonomers
of both types dominate the weight distribution, $\text{w}_{\text{A}},\text{w}_{\text{B}}\rightarrow1/2$.
In the opposite limit of $c_{\text{t}}\gg0$, there is $p\rightarrow1$
and $\omega\rightarrow0$ such that long linear chains dominate leading
to a random distribution of chain ends: $\text{w}_{\text{m}}\rightarrow1/2$
and $\text{w}_{\text{A}},\text{w}_{\text{B}}\rightarrow1/4$.

\begin{figure}
\includegraphics[angle=270,width=1\columnwidth]{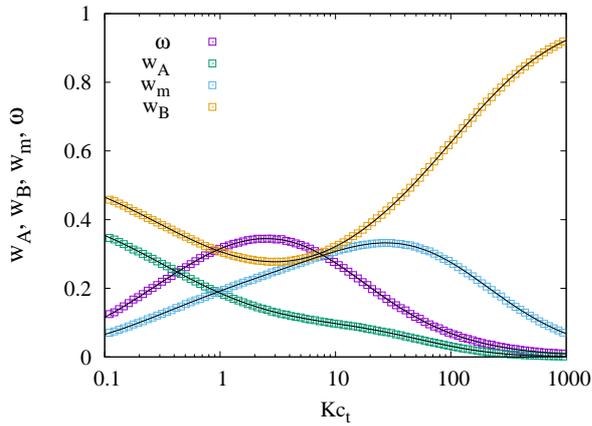}

\caption{\label{fig:Number-fractions-of}Weight fractions of $A$, $B$, and
$m$-terminated chains, and loops for same parameters as in Figure
\ref{fig:Weight-fractions-of} except of $r=0.8$).}
\end{figure}

In Figure \ref{fig:Number-fractions-of}, we show the weight fractions
of $A$, $B$, $m$-terminated chains and loops for the same parameters
as in Figure \ref{fig:Weight-fractions-of} except of a small stoichiometric
imbalance, $r=0.8$. This imbalance lets the majority species of reactive
groups dominate chain termination in the high concentration limit,
$c_{\text{t}}\gg0$, where other chain types are increasingly suppressed.
Since loop formation requires mixed terminated chains, the disappearance
of the latter reduces also the weight fraction of loops.

Let us use the weight fraction of dimers, equation (\ref{eq:omolin}),
as a simple, rough estimate of the location of the maximum weight
fraction of loops through 
\begin{equation}
\frac{\text{d}\text{w}_{\text{m,0,lin}}}{\text{d}p_{\text{lin}}}=0,\label{eq:maximum}
\end{equation}
since the weight fraction of loops is dominated by the smallest loops.
This condition leads to the equation
\begin{equation}
1-2r_{\text{lin}}p_{\text{lin}}-2p_{\text{lin}}+3r_{\text{lin}}p_{\text{lin}}^{2}=0\label{eq:condition}
\end{equation}
where only the negative branch of the solutions
\begin{equation}
p_{\text{opt}}\approx\frac{r_{\text{lin}}+1\pm\sqrt{r_{\text{lin}}^{2}-r_{\text{lin}}+1}}{3r_{\text{lin}}}\label{eq:popt}
\end{equation}
serves as an estimate for the conversion at the maximum amount of
loops, since the positive branch is $>1$ for all $r_{\text{lin}}<1$.
In the example of Figure \ref{fig:Weight-fractions-of} with $r=0.8$,
a maximum weight fraction of $\approx34.4\%$ of loops is obtained
roughly at $c_{\text{t}}\approx c_{\text{i}}/4$, resulting in a conversion
$p\approx0.64$ for $K=10^{3}$. Both $p$ and $\omega$ are clearly
smaller at the maximum as in case 1 polymerization for the same set
of concentrations and reaction constant.

\nomenclature{$p_{\text{opt}}$}{conversion at the maximum amount of loops}%

Let us now compute the number fractions $n_{\text{x}}$ of the different
species inside the full sample. Recall that the number fractions $n_{\text{x}}$
of section \emph{Case 3 without rings} are normalized to unity within
the linear chain fraction. In order to obtain properly normalized
number fractions within the full sample, we consider first the average
DP of the linear chains,
\begin{equation}
N_{\text{n,lin}}=\frac{1+r_{\text{lin}}}{1+r_{\text{lin}}-2r_{\text{lin}}p_{\text{lin}}},\label{eq:Nnlin}
\end{equation}
see equation (\ref{eq:Nn}). The number density of linear chains per
strand is 
\begin{equation}
n_{\text{C}}=\left(1-\omega\right)/N_{\text{n,lin}},\label{eq:nc}
\end{equation}
which we use to compute the number fraction of rings among all molecules,
\begin{equation}
n_{\text{loops}}=\frac{n_{\text{L}}}{n_{\text{C}}+n_{\text{L}}}.\label{eq:ring fraction}
\end{equation}

\nomenclature{$n_{\text{C}}$}{number density of linear chains per strand}%

\nomenclature{$n_{\text{loops}}$}{number fraction of loops among all molecules}%

As mentioned above, the equations in section \emph{Case 3 without
rings} for the linear species can be used after replacing all $p$
and $r$ by $p_{\text{lin}}$ and $r_{\text{lin}}$ respectively,
while all number fractions need to be multiplied by $1-n_{\text{loops}}$
and all weight fractions by $1-\omega$. Thus, our approach provides
exact numerical solutions for all number fractions, weight fractions,
and distributions.

As for case one, see Ref. \citep{Lang2021a}, we can use these results
to obtain sample average quantities that might be useful for an analysis
of the reactions. For instance, both $n_{\text{C}}$ and $n_{\text{L}}$
set up the total density of molecules among the total number of strands,
thus, they are related to the sample average DP via 
\begin{equation}
\frac{1}{N_{\text{n}}}=n_{\text{C}}+n_{\text{L}}\label{eq:Nninvers}
\end{equation}
\[
=\frac{\left(1-\omega\right)\left(1+r_{\text{lin}}-2r_{\text{lin}}p_{\text{lin}}\right)}{1+r_{\text{lin}}}+\frac{\omega}{N_{\text{n,loops}}}.
\]
Similarly, the weight average DP can be obtained by a weighted average
of the four contributions due to rings, and $A$, $B$, and $m$ terminated
chains. The resulting expressions are readily obtained following the
corresponding steps discussed for case 1 in Ref. \citep{Lang2021a},
but they are quite lengthy. Therefore, we omit an explicit treatment
of these equations here. Instead, we show the resulting data in several
Figures.

\begin{figure}
\includegraphics[angle=270,width=1\columnwidth]{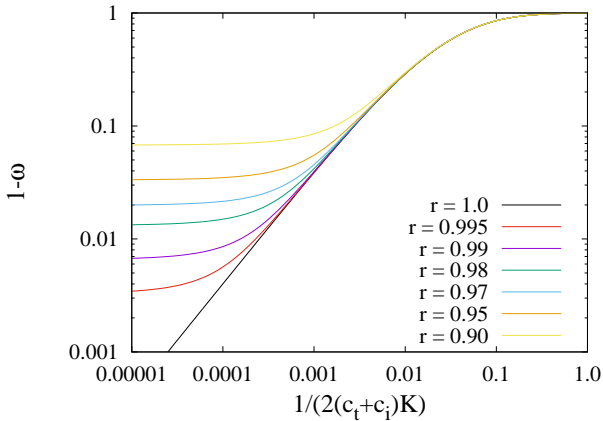}

\caption{\label{fig:Weight-fraction-of-2}Weight fraction of linear chains,
$1-\omega$, as a function of $K$ for $c_{\text{t}}=c_{\text{i}}/10$
and $c_{\text{i}}=1/10$ for stoichimetric ratios close to $r=1$
(case 3).}
\end{figure}

In Figure \ref{fig:Weight-fraction-of-2}, we show the weight fraction
of linear chains, $1-\omega$, as a function of $K$ and dilute concentrations
$c_{\text{t}}=c_{\text{i}}/10$ where the limit $K\rightarrow\infty$
at $r=1$ provides a weight fraction of 100\% rings. As expected,
any small stoichiometric imbalance induces a non-zero weight fraction
of linear chains that is somewhat larger than the lower bound estimate
$\left(1-r\right)/\left(1+r\right)$ (all minority species in rings
and no polymerization of chains) for the weight fraction of linear
chains of in the limit of $K\rightarrow\infty$. Altogether, Figure
\ref{fig:Weight-fraction-of-2} demonstrates that a 100\% weight fraction
of rings is reached only for $r=1$.

In real systems, composition fluctuations arising from the stochastic
motion of the unsaturated reactive groups will control the weight
fraction of rings that can be reached for $r\approx1$ in the limit
of large $K$. For irreversible recombination similar to our case
3, it is well established \citep{Ovchinnikov1978,Toussaint1983,Kang1984}
that these dominate the long time reaction kinetics close to stoichiometrically
balanced conditions, $r\approx1$. A similar dominance of diffusion
and composition fluctuations has been found also for reversible systems
\citep{Rey1999,Gopich2001}, slowing down the convergence towards
complete conversion. Significant diffusion effects are also in conflict
with the assumption of an independence of all reactions, since diffusion
control leads inevitably to higher reaction rates for the faster moving
smaller molecules. Therefore, we expect that the computations of this
section are reasonable approximations for the reaction controlled
limit only up to a limiting $K$ where composition fluctuations or
diffusion effects come into play. Deeper insight into this complex
problem could be obtained with suitable Monte-Carlo or molecular dynamics
simulations as these allow to keep track of the weight dependent mobility
of the molecules including a possible impact of entanglements on polymerization.

One subtle point in this discussion concerns rings having always $r=1$,
so that the linear chains must compensate all of the stoichiometric
imbalance. Any composition fluctuation, thus, reduces both the weight
fraction of rings and the DP of the chains by an amount proportional
to the fluctuating average composition difference in the system. In
our computations, this could be taken into account by replacing the
true $r$ by an effective $r$ that is a function of the particular
reaction constant, since increasing $K$ drives the effective $r$
to unity. However, the details of such a process have not been elaborated
yet for a polymer model system where additional couplings between
the system parameters arise (e.g. the composition fluctuations couple
to the overlap of polymer molecules and the weight fraction of the
rings).

\begin{figure}
\includegraphics[angle=270,width=1\columnwidth]{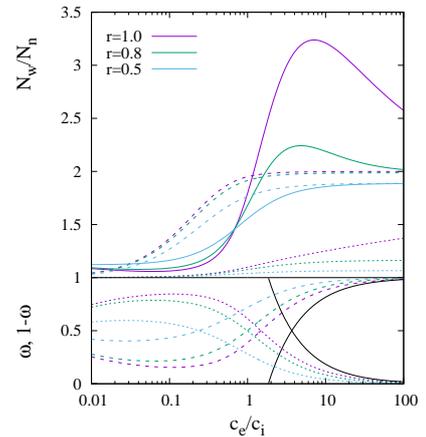}

\caption{\label{fig:Weight-fraction-of}Weight fraction of rings and chains
(lower part) and polydispersity (upper part) at an intermediate reaction
constant $K=10^{4}$ at $c_{\text{i}}=10^{-2}$, for a range of stoichiometric
imbalances $r$. Upper part: the continuous lines address the sample
average polydispersity, the dashed lines the polydispersity of the
dominating (for $r\protect\ne1$) odd chains with end groups of either
type $A$ or $B$, respectively. The dotted lines show the polydispersity
of the rings. Lower part: weight fractions of chains (dashed lines)
and rings (dotted lines). The black line corresponds to the limit
of $K\rightarrow\infty$.}
\end{figure}

The polydispersity of the full sample, of odd numbered chains, and
the rings is shown in Figure \ref{fig:Weight-fraction-of} for a range
of stoichiometric imbalances around $r=1$. The general trend is that
an increasing stoichiometric imbalance limits the average degrees
of polymerization of both linear chains and rings for $c_{\text{t}}\gtrsim c_{\text{i}}$
and thus, it limits the difference in the molecular weights between
these fractions. This leads to a decreasing peak in polydispersity
for a decreasing $r<1$. This trend is reversed for the low concentration
regime, $c_{\text{t}}\lesssim c_{\text{i}}$, since there, increasing
the stoichiometric imbalance is equivalent to introducing larger portions
of non-reacted monomer strands that are shorter than any ring in the
system, which increases polydispersity. Qualitatively similar to case
1, the DP of the chains grows quickly prior to the critical concentration
(polydispersity approaches two), while the polydispersity of the sample
reaches its maximum significantly after the critical concentration.
Thus, a significant number fraction of the linear chains needs to
develop first until a high polydispersity is reached at concentrations
clearly beyond the critical concentration.

For intermediate reaction constants, a peak develops for the weight
fraction of the rings as discussed above, see Figure \ref{fig:Weight-fraction-of}.
This peak turns into a broad plateau in the limit of large $K$. The
Figure shows also the shift of the critical concentration by a factor
of $2^{-1/2}$ as compared to case 1 (see equation (\ref{eq:c_crit}).

\begin{figure}
\includegraphics[angle=270,width=0.8\columnwidth]{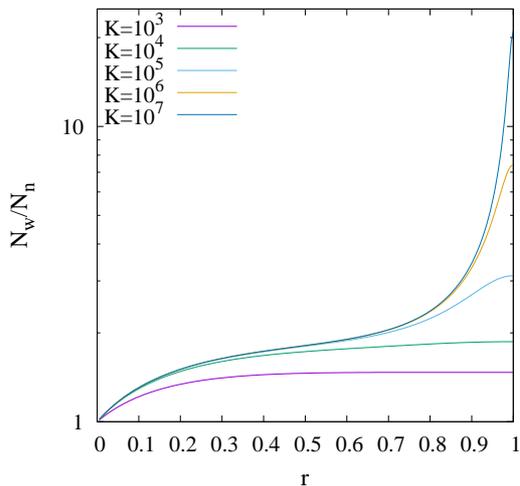}

\caption{\label{fig:Weight-fraction-of-3-1}Polydispersity as a function of
stoichiometric ratio $r$ and reaction constant $K$ (case 3) at $c_{\text{e}}/c_{\text{i}}=4$.}
\end{figure}

With Figure \ref{fig:Weight-fraction-of-3-1}, we compare the impact
of stoichiometric ratio and reaction constant $K$ on the polydispersity.
For this example, we focus on $c_{\text{e}}/c_{\text{i}}=4$, which
is in the range of the polydispersity peaks of Figure \ref{fig:Weight-fraction-of}.
Recall that a high polydispersity requires the coexistence of short
cyclic and long linear molecules at largely different degrees of polymerization
and at a significant weight fraction of the linear chains, which is
why polydispersity is largest for concentrations somewhat larger than
$c_{\text{crit}}$. The main impact of increasing $K$ is to enforce
chain growth, which is not limited by stoichiometric conditions for
$r=1$. For stoichiometric imbalance, $r<1$, there is a particular
reaction constant, where chains are terminated equivalently due to
missing bonds and due to excess majority strands. This point is roughly
visible in Figure \ref{fig:Weight-fraction-of-3-1} by the point where
data for a larger reaction constant than a particular $K$ separate
towards a larger polydispersity, i. e. a larger DP of the linear chains.
At an $r$ below the separation point, the data for the corresponding
reaction constants are dominated by the stoichiometric condition.
At large $r$ close to unity, the impact of a large $K$ stands out
for a broad range of reaction constants and leads to a large increase
of polydispersity (DP of the linear chains) as a function of $K$.

This latter observation could be used to check the preparation conditions
and the quality of the macromonomers, since any impurity, missing,
or inactive group will shift the maximum away from $r=1$, if the
there are more defects in one of the species as compared to the other.
If a similar amount of defects arises within both types of macromonomers,
still the DP of the linear chains might stagnate for an increasing
$K$, which could be traced by analyzing related quantities like,
for instance, the viscosity of the supramolecular solution or melt.

\section{$AB$ monomers with two orthogonal reactions (case 2b) \label{subsec:Case-2-R}}

For case 2b, the same $A-B$ monomers form only bonds between two
$A$ groups or between two $B$ groups, like the supramolecular polymers
of Ref. \citep{Groeger2011}. The weight distributions of case 2b
are similar to the weight distributions of case 3 polymerization for
a stoichiometric ratio $r=1$. However, there are now two different
types of ``dimers'' with either $A$ or $B$ groups on their ends
that can form the smallest possible loops, see Figure \ref{fig:The-three-classical}.
The conversions of these groups, $p_{\text{A}}$ and $p_{\text{B}}$,
may differ significantly, $p_{\text{A}}\ne p_{\text{B}}$. Furthermore,
the concentration of $A$ and $B$ groups is $c_{\text{A}}=c_{\text{B}}=c_{\text{t}}/2$
where $c_{\text{t}}$ is the concentration of all reactive groups.
These deviations from case 3 cause significant quantitative and qualitative
modifications that require an explicit discussion.

\nomenclature{$p_{\text{X}}$}{conversion of $X=A,B$ groups}%

\nomenclature{$p_{\text{X,lin}}$}{conversion of $X=A,B$ groups in chain fraction}%

\nomenclature{$K$}{reaction constant}%

\nomenclature{$K_{\text{X}}$}{reaction constant for $X=A,B$ groups}%

We consider that conversions $p_{\text{A}}$ and $p_{\text{B}}$ of
the $A$ and $B$ groups are independent from each other and given
by the corresponding law of mass action, equation (\ref{eq:px}).
As for case 3, the strands in the loops are at 100\% conversion, and
we have to renormalize the conversions $p_{\text{X}\text{,lin}}$
of the reactive groups $X=A,B$ within the linear chain fraction for
each reactive group separately:
\begin{equation}
p_{\text{X}\text{,lin}}=\frac{p_{\text{X}}-\omega}{1-\omega}.\label{eq:pxlin}
\end{equation}
Similar to case 3, the smallest possible loops are formed from dimers,
however, there are now $A$-terminated dimers and $B$-terminated
dimers that contribute to the formation of the smallest loop. For
$p_{\text{A}}\ne p_{\text{B}}$, we introduce reaction constants $K_{\text{A}}$
and $K_{\text{B}}$ to describe the reversible $A$ and $B$ bonds
respectively, see section \emph{Case 2b without rings} of the Appendix
for more details. Loop formation occurs now either by pairs of $A$
or $B$ bonds, respectively, in balance with the corresponding backwards
reaction.

The weight fraction of the non-reacted monomer is
\begin{equation}
\text{w}_{0}=\text{w}_{\text{m,0}}\left(1-\omega\right)=\left(1-p_{\text{A,lin}}\right)\left(1-p_{\text{B,lin}}\right)\left(1-\omega\right),\label{eq:w0 case2}
\end{equation}
see equation (\ref{eq:omz}) with $z=0$. Here, there are two reactions
with $A$ and $B$ groups respectively, leading to the formation of
strands with one reacted group, $\text{w}_{1}$. For this particular
case, however, we must distinguish these as $\text{w}_{1,\text{A}}$
and $\text{w}_{1,\text{B}}$ where the suffix indicates the type of
the reacted end group. Thus,
\begin{equation}
\text{w}_{\text{1,A}}=c_{\text{A}}\left(1-p_{\text{A}}\right)K_{\text{A}}\text{w}_{0}.\label{eq:w1B}
\end{equation}
\begin{equation}
\text{w}_{\text{1,B}}=c_{\text{B}}\left(1-p_{\text{B}}\right)K_{\text{B}}\text{w}_{0}.\label{eq:w1A}
\end{equation}
\begin{equation}
\text{w}_{\text{1}}=\text{w}_{1,\text{A}}+\text{w}_{1,\text{B}}.\label{eq:w1-2}
\end{equation}
These two types of $\text{w}_{1}$ strands lead to the formation of
$\text{w}_{2}$ strands that are not part of any loop. In analogy
to equation (\ref{eq:w2}), we obtain
\begin{equation}
2\left(\text{w}_{2}-\omega\right)=c_{\text{A}}\left(1-p_{\text{A}}\right)K_{\text{A}}\text{w}_{\text{1,B}}+c_{\text{B}}\left(1-p_{\text{B}}\right)K_{\text{B}}\text{w}_{\text{1,A}}.\label{eq:w2 case 2}
\end{equation}
Note that the above balance equations allow to compute the conversion
of species $X=A,B$ through 
\begin{equation}
p_{\text{X}}=\text{w}_{2}+\text{w}_{1,\text{X}}.\label{eq:pX}
\end{equation}

\nomenclature{w$_{\text{1,X}}$}{weight fraction of strands with one reacted group of type $X=A,B$}%

The weight fraction of the smallest loop, $\omega_{1}$, couples to
the weight fractions of the smallest $A$ and $B$ terminated molecules,
\begin{equation}
\text{w}_{\text{A,0,lin}}=p_{\text{B,lin}}\left(1-p_{\text{A,lin}}\right)^{2}\left(1-\omega\right)/2\label{eq:oalin}
\end{equation}
 and 
\begin{equation}
\text{w}_{\text{B,0,lin}}=p_{\text{A,lin}}\left(1-p_{\text{B,lin}}\right)^{2}\left(1-\omega\right)/2\label{eq:oblin}
\end{equation}
respectively, see equation (\ref{eq:oaz}) and (\ref{eq:obz}) and
Figure \ref{fig:The-three-classical}. The smallest loop is a dimer
where two bonds can break. This leads to the balance equation 
\begin{equation}
2\omega_{1}=2^{-3/2}c_{\text{i}}\left(K_{\text{A}}\text{w}_{\text{A,0,lin}}+K_{\text{B}}\text{w}_{\text{B,0,lin}}\right)\label{eq:weins}
\end{equation}
for the smallest loop and a total weight fraction of loops of
\begin{equation}
\omega=\omega_{1}\sum_{z=1}^{\infty}z^{-3/2}\left(p_{\text{A,lin}}p_{\text{B,lin}}\right)^{z-1}.\label{eq:loops case 2}
\end{equation}
In total, we obtain in the limit of $c_{\text{t}}\gg c_{\text{i}}$
and $K_{\text{X}}\rightarrow\infty$ where $p_{\text{A,lin}}p_{\text{B,lin}}\rightarrow1$
that $\omega$ remains smaller than in case 1 by a factor of $2^{1/2}$
similar to case 3 for $r=1$. As before, the above set of equations
can be solved exactly using the numerical scheme discussed in the
Appendix of Ref. \citep{Lang2021a}.

With these equations solved, we proceed to the computation of the
distribution functions and averages. Here, we require the number density
of loops per initial strand, 
\begin{equation}
n_{\text{L}}=\frac{\omega_{1}}{2}\sum_{z=1}^{\infty}z^{-5/2}\left(p_{\text{A,lin}}p_{\text{B,lin}}\right)^{z-1},\label{eq:enel}
\end{equation}
and the average DP of the loops that is computed using equation (\ref{eq:Nnloops}).
The factor $1/2$ in equation (\ref{eq:enel}) takes into account
that loops are formed by pairs (or multiple pairs) of $A-B$ strands.
The average DP of the linear chains is given by equation (\ref{eq:Nn-2})
after replacing $p_{\text{A}}$ and $p_{\text{B}}$ with the corresponding
expressions of the linear chain fraction, $p_{\text{A,lin}}$ and
$p_{\text{B,lin}}$. Finally, the number density of chains per strand,
$n_{\text{C}}$, and the number fraction of loops among all molecules,
$n_{\text{loops}}$, are computed with equation (\ref{eq:nc}) and
(\ref{eq:ring fraction}).

After these results have been obtained, the number and weight fractions
of the linear chains of section \emph{Case 2b without rings} must
be normalized by a factor $1-n_{\text{loops}}$ and $1-\omega$, respectively,
to reflect the corresponding contributions to the full sample similar
to the preceding section. Here, again $p_{\text{A,lin}}$ and $p_{\text{B,lin}}$
replace $p_{\text{A}}$ and $p_{\text{B}}$ in all expressions. Either
through the resulting number and weight distributions or by combining
the corresponding averages as we have done in the preceding section
and in Ref. \citep{Lang2021a}, the sample wide number average and
weight average DP becomes available. We do not provide explicit equations
here, since the corresponding steps have been discussed previously,
the derivation is straightforward, and the resulting expressions are
quite lengthy. As before, we provide Figures with the resulting data
for a range of reaction constants and concentrations.

\begin{figure}
\includegraphics[angle=270,width=1\columnwidth]{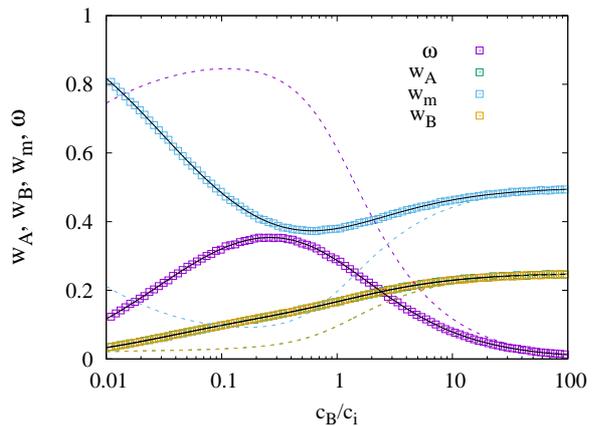}

\caption{\label{fig:Weight-fractions-r1}Weight fractions $\text{w}_{\text{A}}$,
$\text{w}_{\text{B}}$, $\text{w}_{\text{m}}$, and $\omega$ for
$K_{\text{A}}=K_{\text{B}}=100$ (symbols for simulation data and
continuous lines for theory) and $c_{\text{i}}=10^{-2}$ (case 2b).
Dashed lines represent theory for $K_{\text{A}}=K_{\text{B}}=1000$
to indicate the trend as a function of increasing $K$.}
\end{figure}

\begin{figure}
\includegraphics[angle=270,width=1\columnwidth]{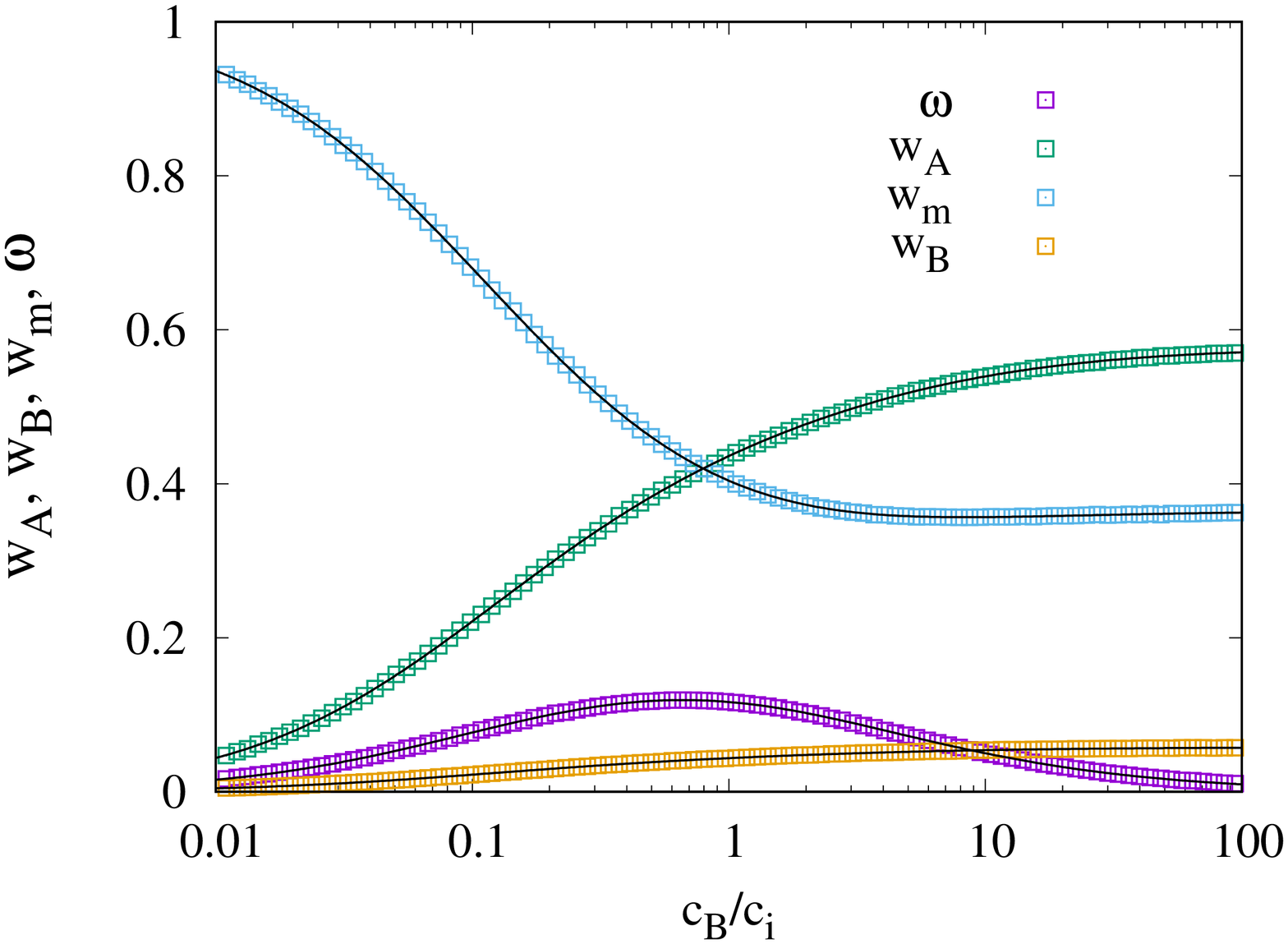}

\caption{\label{fig:Weight-fractions-,}Weight fractions $\text{w}_{\text{A}}$,
$\text{w}_{\text{B}}$, $\text{w}_{\text{m}}$, and $\omega$ for
$K_{\text{A}}=100$, $K_{\text{B}}=1000$, and $c_{\text{i}}=10^{-2}$
(case 2b).}
\end{figure}

We have tested our equations by comparing the exact numerical solution
of the balance equations with Monte Carlo simulation data. Figure
\ref{fig:Weight-fractions-r1} provides the corresponding data for
the weight fractions of $A$, $B$, and $m$-terminated chains of
symmetric cases at different reaction constants. Similar to case 3,
the limiting case of $c_{\text{B}}\rightarrow0$ provides the distributions
of the macromonomers without reactions, which is $\text{w}_{\text{m}}\rightarrow1$,
while all other contributions decay to zero. Again, the limit of $Kc_{\text{B}}\rightarrow\infty$
produces no rings but infinitely long chains with a random distribution
of ends $\text{w}_{\text{m}}\rightarrow1/2$, and $\text{w}_{\text{A}},\text{w}_{\text{B}}\rightarrow1/4$.
Between these limits, a significant or even dominant portion of rings
is being formed that increases with increasing reaction constant $K$.
As for case 3, the optimum conditions for ring formation can be estimated
by analyzing the maximum of the dimers, which contains here two contributions
and is more complex to analyze. For the sake of brevity, we omit an
explicit discussion here and mention that this peak turns into a broad
plateaux with $\omega\approx1$ for $c_{\text{B}}<c_{\text{i}}$ and
$c_{\text{B}}K\gg1$ in similar manner as for case 2b.

Figure \ref{fig:Weight-fractions-,} shows data for an asymmetric
case with a smaller reaction constant $K_{\text{A}}$. Here $A$ groups
predominantly terminate chains, once these grow for $Kc_{\text{A}}>1$.
This reduces significantly the fraction of loops together with the
DP of the linear chains and loops.

\begin{figure}
\includegraphics[angle=270,width=1\columnwidth]{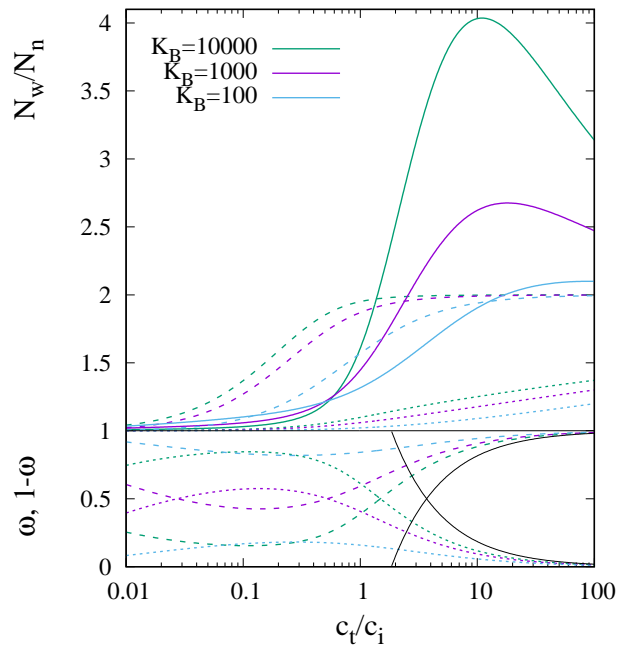}

\caption{\label{fig:Weight-fractions-of-1}Weight fractions of loops and polydispersities
for $K_{\text{A}}=10^{4}$ and $c_{\text{i}}=10^{-2}$ as a function
of the total concentration of reactive groups, $c_{\text{t}}$, for
different $K_{\text{B}}$ (case 2b).}
\end{figure}

In Figure \ref{fig:Weight-fractions-of-1}, we show the weight fractions
of loops and chains (bottom) along with the polydispersity of the
sample for a range of different $K_{\text{B}}$ at a fixed $K_{\text{A}}$.
The qualitative behavior is quite similar to case 3 at a significantly
reduced weight fraction of loops and degrees of polymerization similar
to the preceding Figure. The polydispersity peak of the sample is
reached at significantly larger concentration as the critical concentration.
The critical concentration itself is shifted by a factor of $2^{1/2}$
to lower concentrations as compared to case 1, again in accord with
case 3 for $r=1$. Another interesting point is that the weight distributions
of this case are similar to $r=1$ of case 3. However, composition
fluctuations of reactive groups are largely suppressed here, and they
couple to the concentration fluctuations of the polymers on large
length scales.

\section{Discussion\label{sec:Discussion}}

\begin{figure}
\includegraphics[angle=270,width=1\columnwidth]{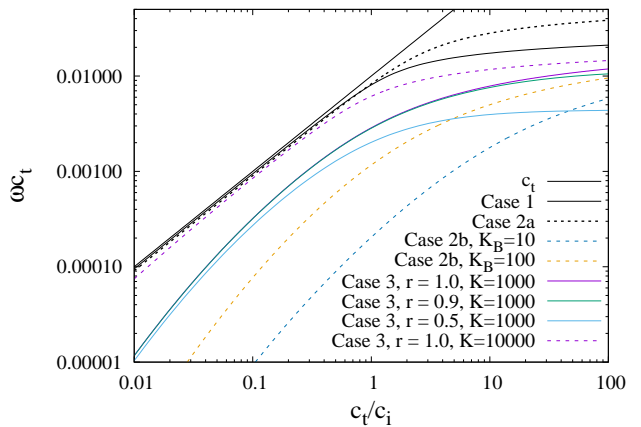}

\caption{\label{fig:Effective-molarity-for}Weight fraction of loops in total
sample including a solvent of same density for different model systems
with $K=K_{\text{A}}=10^{3}$ (unless indicated otherwise) and $c_{\text{i}}=10^{-2}$.}
\end{figure}

Let us start our with a quantitative comparison of loop formation
in all cases discussed in the preceding sections and in Ref. \citep{Lang2021a}.
This comparison is shown in Figure \ref{fig:Effective-molarity-for}
using the same representation of the data as in the work of Ercolani
\citep{Ercolani1993} for a better comparison with preceding work.
For our discussion, we have also compiled the balance equations and
critical concentrations in table \ref{tab:Short-representation-of}.
In Figure \ref{fig:Effective-molarity-for}, we have multiplied the
weight fraction of loops with the total concentration of reactive
groups, $c_{\text{t}}$, which is proportional to the concentration
of macromonomers. If the solvent has the same density as the macromonomer,
the $y$-axis is equivalent to the weight fraction of rings in the
total sample. For comparison, $c_{\text{t}}$ is included referring
to $\omega=1$. In the low concentration regime, $c_{\text{t}}<c_{\text{i}}$,
the weight fraction of loops is essentially the weight fraction of
the macromonomers for case 1 and case 2a, settling to an almost constant
amount of rings around the critical concentration. As discussed above,
the cross-over occurs later for case 2a, leading to about twice as
much rings in the high concentration limit. Small differences between
case 1 and 2 in the low concentration limit result from a different
conversion because either $K$ or $2K$ enters in the law of mass
action. The Figure contains also one example of case 3 at $r=1$ and
a ``high'' reaction constant $K=10^{4}$. This cases settles at
a lower amount of loops by a factor of $2^{-1/2}$, which stems from
a lower critical concentration, see Table \ref{tab:Short-representation-of}.
Thus, the dependence of the uppermost three sets of data demonstrates
that nature prefers to make loops in the large $K$ limit for concentrations
up to $c_{\text{crit}}$, while for $c>c_{\text{crit}}$, the excess
concentration of macromonomers beyond $c_{\text{crit}}$ is predominantly
converted into linear chains.

\begin{table*}
\begin{center}%
\begin{tabular}{|c|c|c|c|c|}
\hline 
 & case 1 & case 2a & case 2b & case 3\tabularnewline
\hline 
\hline 
$w_{1}$ & $4\left(1-p\right)c_{\text{t}}K\text{w}_{0}$ & $\left(1-p\right)c_{\text{t}}K\text{w}_{0}$ & $\left[\left(1-p_{\text{A}}\right)c_{\text{A}}K_{\text{A}}+\left(1-p_{\text{B}}\right)c_{\text{B}}K_{\text{B}}\right]\text{w}_{0}$ & $2\left(1-rp\right)\frac{c_{\text{t}}K}{r+1}\text{w}_{0}$\tabularnewline
\hline 
$\text{w}_{2}$ & $4\left[\left(1-p\right)c_{\text{t}}K\right]^{2}\text{w}_{0}+\omega$ & $\left[\left(1-p\right)c_{\text{t}}K\right]^{2}\text{w}_{0}/4+\omega$ & $\left(1-p_{\text{A}}\right)\left(1-p_{\text{B}}\right)c_{\text{A}}c_{\text{B}}K_{\text{A}}K_{\text{B}}\text{w}_{0}+\omega$ & $\left[\left(1-rp\right)\frac{c_{\text{t}}K}{\left(r+1\right)}\right]^{2}\text{w}_{\text{0}}+\omega_{\text{A}}$\tabularnewline
\hline 
$\omega$ & $2c_{\text{i}}K\text{w}_{0}\sum_{k}p_{\text{lin}}^{k-1}$ & $c_{\text{i}}K\text{w}_{0}\sum_{k}p_{\text{lin}}^{k-1}$ & $2^{-5/2}c_{\text{i}}\left(K_{\text{A}}\text{w}_{\text{A,0,lin}}+K_{\text{B}}\text{w}_{\text{B,0,lin}}\right)\sum_{k}q^{k-1}$ & $\frac{2^{-5/2}}{1+r}c_{\text{i}}K\text{w}_{\text{m,0,lin}}\sum_{k}q^{k-1}$\tabularnewline
\hline 
$c_{\text{crit}}/c_{\text{i}}$ & $\sum_{k}$ & $2\sum_{k}$ & $2^{-1/2}\sum_{k}$ & $2^{-1/2}\sum_{k}$ for $r=1$\tabularnewline
\hline 
equations & (13)-(15), (29), (30) of \citep{Lang2021a} & - & (6), (7), (9), (12) & (25)-(28), (32), (33)\tabularnewline
\hline 
\end{tabular}

\end{center}

\caption{\label{tab:Short-representation-of}Short representation of balance
equations (example: $w_{1}=4\left(1-p\right)c_{\text{t}}K\text{w}_{0}$
for case 1) for $\text{w}_{1}$, $\text{w}_{2}$ and $\omega$ and
the critical concentration. Equations for case 2b refer to reactions
of $A$ groups only. The shorthand $\sum_{k}=\sum_{k=1}^{\infty}k^{-3/2}$
is used for all cases, while $q$ abbreviates $r_{\text{lin}}p_{\text{lin}}^{2}$
in case 2b and $p_{\text{A,lin}}p_{\text{B,lin}}$ in case 3.}
\end{table*}

For case 3 and case 2b, the weight fraction of loops is typically
smaller than the total weight fraction of macromonomers in the low
concentration regime for intermediate values of $K$ or $r<1$, and
it must not increase linearly as shown by the data. Only for very
large reaction constants and nearly stoichiometric conditions, the
weight fraction of rings approaches the weight fraction of the macromonomers
in the low concentration regime. Correspondingly, in the high concentration
limit, the weight fraction of rings settles at lower total amounts.
Off-stoichiometric conditions or a second lower reaction constant
reduce sifgnificantly the portion of rings in these cases. Thus, if
a high yield of rings is desired, case 2a is the best choice. If loop
formation should be suppressed, case 2b is preferable, since it avoids
problems related to composition fluctuations and stoichiometric balance
that might arise in case 3.

The differences in the mathematical description of the four examples
are highlighted in Table \ref{tab:Short-representation-of}. Case
2a is - within our mean-field treatment - equivalent to case 1 except
of factors of 2 regarding $K$ and $c_{\text{t}}$. The main difference
between case 2b and the reference case 1 is that there are two distinct
reaction mechanisms that lead to bond formation, which show up in
separate contributions for $\text{w}_{1}$ and $\omega$, while $\text{w}_{2}$
results from a combination of both mechanisms. The numerical coefficient
for $\omega$ highlights that loops are formed from pairs of macrmonomers,
while $c_{\text{i}}$ is defined with respect to end-contacts of macromonomers.
Case 3 turns into case 2a for $r=1$ regarding linear chain growth,
which regards $\text{w}_{1}$ and $\text{w}_{2}$, while the difference
for $\omega$ reflects that loops are formed from dimers. The probability
$q$ for adding a dimer to a linear chain inside the linear fraction
(case 2b and 3) takes over the role of $p_{\text{lin}}$ in case 1
and 2a, where the latter is the probability to add another monomer
within the linear chain fraction there.

\begin{figure}
\includegraphics[angle=270,width=1\columnwidth]{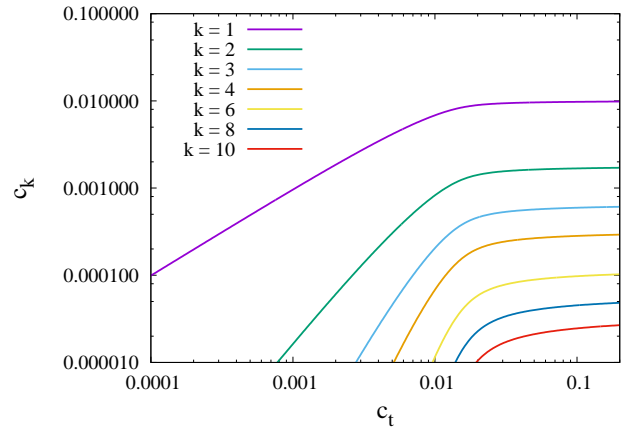}

\caption{\label{fig:Effective-molarity-for-1}Concentration of loops made of
$k$ strands as a function of $c_{\text{t}}$ for $K=10^{4}$ and
$c_{\text{i}}=10^{-2}$ for case 1.}
\end{figure}

\noindent Figure \ref{fig:Effective-molarity-for-1} shows the concentration
of loops made of $k$ strands as a function of $c_{\text{t}}$ for
case 1. The data in this plot is presented in similar form as the
data in Ref. \citep{Ercolani1993} and refers to unstrained ring polymers
where no additional entropic or energetic penalty applies for the
smallest molecules. We have included this plot to demonstrate that
our computation is fully equivalent to preceding work based upon Ref.
\citep{Ercolani1993}. A treatment of strained rings and other corrections
regarding the formation of rings (e.g. bond correlations, etc.) can
be considered by the summation over all cyclic states for determining
the weight fraction of loops (equation (29) of Ref. \citep{Lang2021a},
(\ref{eq:omega}), and (\ref{eq:loops case 2}) correspondingly).
An excellent guide to these corrections is the recent review by Di
Stefano and Mandolini \citep{DiStefano2019}.

Recent literature provides a plethora of examples for supramolecular
self-assembly where more than two compounds or types of bonds are
involved. Winter and Schubert, for instance, distinguish six different
classes of supramolecular polymerization only regarding metallo-supramolecular
polymers, see Figure 3 of Ref. \citep{Winter2016}. Here class Ia)
is equivalent to case 1, Ib) and IIa) are case 3 in our notation,
and Ic is case 2a. Molecular weight distributions for more complex
classes like IIb) and IIc) of Ref. \citep{Winter2016} can be derived
using our results. For instance, class IIc) is equivalent to case
3 after considering that loop formation is enhanced by a factor of
two similar to case 2a, while class IIb) requires the consideration
of quadruples of units with two instead of one stoichiometric ratio.
Thus, class IIb) is also a generalization of case 3. Our theoretical
analysis may serve as a template to derive weight distributions for
these and other more complex cases.

In the sections above, we have discussed only LGS which refers to
polymer melts or $\theta$-solutions. A generalization to good or
a-thermal solvent is discussed in Ref. \citep{Lang2021a} along with
a discussion of poor solvents that applies also to case 2b where all
macromonomers are of the same type. Case three in the presence of
a poor and probably selective solvent with reactions occurring across
a phase boundary is a quite complex problem that goes clearly beyond
the scope of the present paper. We postpone the discussion of this
subject to forthcoming work.

A generalization with respect to a first shell substitution effect
is straight forward, since this effect requires to consider different
reaction constants in the two balance equations that connect either
$\text{w}_{1}$ with $\text{w}_{0}$ and $\text{w}_{1}$ with $\text{w}_{2}.$
Similarly, the balance equations of loops need to be modified, if
loop closure occurs starting from a chain end on $\text{w}_{1}$ or
on $\text{w}_{2}$. More details on the implementation of a first
shell substitution effect can be found in our preceding work \citep{Lang2021a}.

Our approach provides molecular weight distributions for linear chains
and rings in theta solvents (good solvents require some adjustments
discussed in Ref. \citep{Lang2021a} and could be computed numerically).
These can be incorporated in models for the properties of the corresponding
supramolecular solutions that do consider polydispersity. We expect
that such a generalization provides a more accurate analsis of experimental
data of supramolecular solutions.

\section{Summary}

In the present work, we have developed an exact numerical solution
for the stepwise reversible alternating co-polymerization of two strands
of type $A$ and $B$ and the stepwise reversible polymerization of
linear precursors where both ends participate in two orthogonal reactions.
Both systems were treated exactly in the mean field limit for both
cases with and without cyclization. Our discussion shows that the
system with the orthogonal reactions is particularly suited to suppress
cyclization in contrast to a reaction of the same chain architecture
where the ends undergo a hetero-complementary coupling of $A$ with
$B$ reactive groups. This latter system leads to largest weight fractions
of cyclics at otherwise identical parameters like intra- and inter-molecular
concentrations of reactive groups and an identical reaction constant.

All four systems that we have studied develop a comparatively large
polydispersity (see also Ref. \citep{Lang2021a} regarding case 1
and case 2a at concentrations about four to ten times larger than
the intra-molecular concentration. The critical concentration is not
universal, it appears at different ratios of the inter- to the intra-molecular
concentration of the reactive groups depending on the particular reaction
mechanism. One important point of the discussion is that cyclic species
are always at 100\% conversion and are always stoichiometrically balanced
in case of an alternating co-polymerization. Therefore, any deviation
from stoichiometry or complete reactions is compensated by the remaining
linear species alone. This causes a systematic split of the properties
of the linear and cyclic chain fractions and shows a strong impact
on the corresponding distributions and averages.

We have tested our theory by comparing with Monte Carlo simulations
that were developed in Ref. \citep{Lang2021a}. These simulations
resemble the mean field limit by employing a set of Gaussian strands
that react only according to given concentrations and reaction constants.
The observed excellent agreement with the simulation data, therefore,
is a strong support for our analytical discussion. We expect that
our work will be applied to develop theory for more complex supramolecular
systems and regarding a more accuarate analysis of experimental data.

\section{Acknowledgements}

The authors thank the ZIH Dresden for a generous grant of computation
time and the DFG for funding Project LA2735/5-1. The authors also
thank Frank Böhme, Reinhard Scholz, and Toni Müller for useful comments
on earlier versions of the manuscript.

\section{Appendix}

\setcounter{equation}{0}

\renewcommand{\theequation}{A\arabic{equation}}

\subsection{\label{subsec:Weight-distributions-and}Case 3 without rings}

As mentioned in the main part of this work, we compute number and
weight distributions of all classes of chains, specific averages and
total average degrees of polymerization, since previous work contains
some obvious mistakes (wrong normalization \citep{Flory1936}, non-integer
powers of some probabilities \citep{Mizerovskii2013}, etc.). Furthermore,
not all required distributions and averages were computed before.
This gap is closed with the derivation below.

We consider a case 3 polymerization where LGS with functional groups
of type $A$ react exclusively with functional groups of type $B$
that are located on a second fraction of LGS. The molar ratio of type
$A$ strands to type $B$ strands is defined as the stoichiometric
ratio $r$, see equation (\ref{eq:r}). For simplification, we assume
that both strands have roughly the same molar mass, which allows a
simplified treatment based upon degrees of polymerization that we
define with respect to the number of precursor strands in one molecule.
As notation, we use $p_{\text{A}}=p$ for the conversion of the minority
$A$ groups, $c_{\text{A}}$ and $c_{\text{B}}$ for the concentration
of $A$ and $B$ groups respectively, and $c_{\text{o,A}}$ and $c_{\text{o,B}}$
as the concentration of the non-paired $A$ and $B$ groups respectively.

\nomenclature{$c_{\text{o,X}}$}{concentration of non-paired groups of type $X=A,B$}%

We introduce the total concentration of reactive groups, $c_{\text{t}}$,
as the sum of the total concentrations of $A$ and $B$ groups, see
equation (\ref{eq:ct}). Since total conversion is defined with respect
to the maximum possible conversion, there is 
\begin{equation}
p=p_{\text{A}}=1-\frac{c_{\text{o,A}}}{c_{\text{A}}}\label{eq:ppa}
\end{equation}
and the conversion of the $B$ groups is
\begin{equation}
p_{\text{B}}=rp=1-\frac{c_{\text{o,B}}}{c_{\text{B}}}.\label{eq:prb}
\end{equation}

The concentration of the reaction products is the concentration of
the reacted minority $A$ groups, $c_{\text{A}}-c_{\text{o,A}},$
since there is one reacted $A$ group per bond. Note that this concentration
equals the concentration of the reacted $B$ groups, 
\begin{equation}
c_{\text{B}}-c_{\text{o,B}}=c_{\text{A}}-c_{\text{o,A}}\label{eq:cdiff}
\end{equation}
and that the concentration of the open $A$ groups can be expressed
in terms of concentrations of the $B$ groups
\begin{equation}
c_{\text{o,A}}=c_{\text{o,B}}-c_{\text{B}}+c_{\text{A}}=c_{\text{o,B}}+c_{\text{B}}\left(r-1\right).\label{eq:coA}
\end{equation}
The total concentration of reactive groups is 
\begin{equation}
c_{\text{A}}+c_{\text{B}}=\left(r+1\right)c_{\text{B}}.\label{eq:ctot}
\end{equation}
Thus, the product of the concentration of the reactants is 
\begin{equation}
c_{\text{o,A}}c_{\text{o,B}}=c_{\text{o,B}}\left(c_{\text{o,B}}+c_{\text{B}}\left(r-1\right)\right)\label{eq:coAB}
\end{equation}
This leads to a law of mass action with reaction constant
\begin{equation}
K=\frac{c_{\text{B}}-c_{\text{o,B}}}{c_{\text{o,B}}\left(c_{\text{o,B}}+c_{\text{B}}\left(r-1\right)\right)}.\label{eq:K-1}
\end{equation}
Note that there is exactly one $A$ and $B$ group per bond, thus,
there is no factor of two in the definition of $K$ in contrast to
case 1, see Ref. \citep{Lang2021a}. This last equation can be solved
for $c_{\text{o,B}}$, which provides

\begin{equation}
c_{\text{o,B}}=\label{eq:coB}
\end{equation}
\[
\frac{\left(c_{\text{B}}^{2}K^{2}\left(r-1\right)^{2}+2c_{\text{B}}K(r+1)+1\right)^{1/2}+c_{\text{B}}K(1-r)-1}{2K}
\]
and conversion of $B$ groups through
\begin{equation}
p_{\text{B}}=1-\frac{c_{\text{o,B}}}{c_{\text{B}}}\label{eq:pB}
\end{equation}
Finally, $p$ is given by $p_{\text{B}}/r$, see equation (\ref{eq:prb}).

\begin{figure}
\includegraphics[angle=270,width=1\columnwidth]{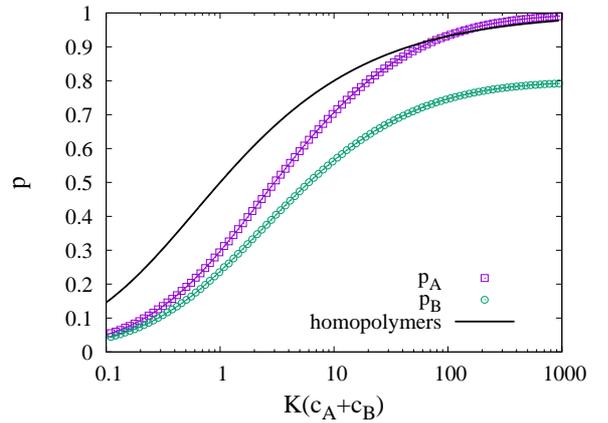}

\caption{\label{fig:-and--1}$p_{\text{A}}$ and $p_{\text{B}}$ as a function
of $\left(c_{\text{A}}+c_{\text{B}}\right)K$ for a stoichiometric
$r=0.8$ in the absence of loop formation, $c_{\text{i}}=0$ (case
3). The thick black line refers to reversible homopolymers with identical
end groups on both chain ends, equation (4) of Ref. \citep{Lang2021a}.}
\end{figure}

This solution is tested with simulation data in Figure \ref{fig:-and--1},
and excellent agreement is found. Note that the above solution does
not converge towards equation (4) of Ref. \citep{Lang2021a} for the
same total concentration of reactive groups, since $c_{\text{A}}=c_{\text{B}}$
for $r=1$. In fact, the case $r=1$ is equivalent to the homopolymerization
of $A-B$ monomers (case 2a) where $A$ groups react exclusively with
$B$ groups. Note that $p_{\text{A}}$ is significantly below one
for a broad range $Kc_{\text{B}}$. Therefore, the 100\% case discussed
in Ref. \citep{Jacobson1950} with $p_{\text{A}}=1$ serves as an
approximation only in the limit of large $Kc_{\text{B}}$.

\nomenclature{$P_{\text{end}(A)}$}{fraction of non-reacted $A$ groups among all chain ends}%

In the reaction product, we classify the resulting molecules whether
these are terminated by only $A$ groups, only $B$ groups or both
(mixed ``$m$'' terminated). The weight fractions of these different
types of chains will be denoted below by $\text{w}_{\text{A}}$, $\text{w}_{\text{B}}$,
and $\text{w}_{\text{m}}$ respectively, while the number fractions
are denoted by $n_{\text{A}}$, $n_{\text{B}}$, and $n_{\text{m}}$.
For the computation of the molecular weight fractions, we start with
the total density of chain ends that is given by the total concentration
of non-paired (``open'') reactive groups 
\begin{equation}
c_{\text{o}}=c_{\text{A}}\left(1-p\right)+c_{\text{B}}\left(1-rp\right)=\left(1+r-2rp\right)c_{\text{B}}.\label{eq:cnon}
\end{equation}
The ratio between the concentration of all reactive groups and $c_{\text{o}}$
provides the average DP 
\begin{equation}
N_{\text{n}}=\frac{c_{\text{A}}+c_{\text{B}}}{c_{\text{o}}}=\frac{1+r}{1+r-2rp}.\label{eq:Nn}
\end{equation}
For a simple derivation of the number fraction distributions, let
us introduce an unknown normalization constant $Y$ to be determined
later. The probability that a chain end of type $A$ is selected randomly
as a starting point of a chain, $P_{\text{end}}\left(A\right)$ is
equal to the portion of non-reacted $A$ groups among all non-reacted
chain ends, 
\begin{equation}
P_{\text{end}}\left(A\right)=\frac{c_{\text{A}}\left(1-p\right)}{c_{\text{o}}}=\frac{r(1-p)}{1+r-2rp}.\label{eq:end_A}
\end{equation}
Furthermore, a portion of $(1-p)$ of all $A$ groups terminates a
chain. Thus, we expect a dependence for the number fraction of all
$A$ terminated chains like 
\begin{equation}
n_{\text{A}}=P_{\text{end}}\left(A\right)(1-p)X=\frac{r\left(1-p\right)^{2}}{1+r-2rp}Y.\label{eq:na}
\end{equation}
Similarly, $B$ strands are selected as starting points of a chain
with a probability $1-P_{\text{end}}\left(A\right)$ among all chain
ends, while also a portion of $1-rp$ strands terminates a chain.
Thus, 
\begin{equation}
n_{\text{B}}=\left(1-P_{\text{end}}\left(A\right)\right)(1-rp)Y=\frac{\left(1-rp\right)^{2}}{1+r-2rp}Y.\label{eq:nB}
\end{equation}
Finally, mixed terminated chains start from an AB pair of strands.
This pair requires a bond in between that exists with probability
$rp$ (when starting from a $B$ end with probability $1-P_{\text{end}}\left(A\right)$
and terminating at an $A$ with termination probability $1-p$) or
with probability $p$ (when starting from an $A$ end with probability
$P_{\text{end}}\left(A\right)$ and terminating at a $B$ with termination
probability $1-rp$). Together these two cases provide
\begin{equation}
n_{\text{m}}=\frac{2rp(1-p)(1-rp)}{1+r-2rp}Y.\label{eq:nm}
\end{equation}
The $Y$ in the above three cases is a normalization factor that accounts
for the weight distribution of strands. This normalization factor
is conveniently computed from the normalization of the number fractions
\begin{equation}
n_{\text{m}}+n_{\text{A}}+n_{\text{B}}=1.\label{eq:norm1}
\end{equation}
This provides
\[
Y^{-1}=\frac{\left(1-p\right)^{2}r+\left(1-rp\right)^{2}+2rp(1-p)(1-rp)}{\left(r(1-p)+1-rp\right)}
\]
\begin{equation}
=1-rp^{2}\label{eq:X-1}
\end{equation}
for $n_{\text{A}}$, $n_{\text{B}}$, and $n_{\text{m}}$. To understand
the physical origin of $Y$ let us introduce the probability $q=rp^{2}$
that an $AB$ strand pair is added to an existing chain. What we have
left out in our derivation above is the distribution of additional
pairs of $A$ and $B$ strands that are attached to a given set of
chain ends. The probability of finding chain ends with $z$ added
$AB$ pairs decays as $q^{z}.$ Since 
\begin{equation}
\sum_{z=0}^{\infty}q^{z}=\frac{1}{1-q}=\frac{1}{1-rp^{2}}=Y,\label{eq:sumz}
\end{equation}
we see indeed that the normalization factor is the sum over the number
fraction distribution of chains (with the same ends). Note that with
the last two equations we also have demonstrated that the number fraction
distributions are properly normalized.

\nomenclature{$Y$}{normation factor of number fraction distributions}%

\nomenclature{$n_{\text{X,z}}$}{number fraction of $X=A,B,m$ terminated chains with $z$ additional pairs of strands beyond the smallest chains of this type}%

\nomenclature{w$_{\text{X,z}}$}{weight fraction of $X=A,B,m$ terminated chains with $z$ additional pairs of strands beyond the smallest chains of this type}%

Putting together the above relations, we can write down the number
fraction distributions of all chains depending on end groups and the
number $z\ge0$ of additional pairs $AB$ of strands,
\begin{equation}
n_{\text{m,z}}=p^{2z}r^{z}\frac{2rp\left(1-p\right)\left(1-rp\right)}{1+r-2rp}\label{eq:nmz}
\end{equation}
\begin{equation}
n_{\text{A,z}}=p^{2z}r^{z}\frac{r\left(1-p\right)^{2}}{1+r-2rp}\label{eq:nAz}
\end{equation}
\begin{equation}
n_{\text{B,z}}=p^{2z}r^{z}\frac{\left(1-rp\right)^{2}}{1+r-2rp}.\label{eq:nBz}
\end{equation}

The shortest realizations of an $A$, $B$, or $m$-terminated chain
($z=0$) are a single $A$ or $B$ strand or a single $AB$ dimer
respectively. Note that the distributions, equation (\ref{eq:nmz})
to equation (\ref{eq:nBz}), agree with older work (e.g. equation
(20), (29), and (30) of Ref. \citep{Flory1936}). More recent work
arrives at different results (e.g. the fourth equation from the bottom
on page 344 of Ref. \citep{Mizerovskii2013}).

\nomenclature{$y_{\text{X}}(z)$}{function for relating weight with number fractions}%

For simplicity, let us assume that the molar mass of an $A$ strand
equals the molar mass of a $B$ strand. The weight fractions of $A$,
$B$, or $m$ terminated chains are then obtained in the standard
way by multiplying the corresponding number fraction distribution
with the number of strands over $N_{\text{n}}$. For $X=A,B,m$ we
write this as 
\begin{equation}
\text{w}_{\text{X,z}}=\frac{y_{X}(z)}{N_{\text{n}}}n_{\text{X,z}}\label{eq:omega}
\end{equation}
where we use a function 
\begin{equation}
y_{\text{m}}(z)=2\left(z+1\right),\label{eq:yx}
\end{equation}
for mixed terminated chains that becomes
\begin{equation}
y_{\text{X}}(z)=2z+1\label{eq:yx2}
\end{equation}
for $X=A,B$. This function counts the number of strands in a chain
with $z$ additional pairs of strands beyond the smallest chain in
this class.

We obtain 
\begin{equation}
\text{w}_{\text{m,z}}=2\left(z+1\right)p^{2z}r^{z}\left[\frac{2pr\left(1-p\right)\left(1-rp\right)}{1+r}\right]\label{eq:omegam}
\end{equation}
\begin{equation}
\text{w}_{\text{A,z}}=\left(2z+1\right)p^{2z}r^{z}\left[\frac{r\left(1-p\right)^{2}}{1+r}\right]\label{eq:omegaa}
\end{equation}
\begin{equation}
\text{w}_{\text{B,z}}=\left(2z+1\right)p^{2z}r^{z}\left[\frac{\left(1-rp\right)^{2}}{1+r}\right].\label{eq:omegab}
\end{equation}
Note that these equations agree with Flory's work except for $w_{\text{A,z}}$
that contains one extra power in $r$ in comparison with equation
(27) of Ref. \citep{Flory1936}. Most likely, this was just a misprint,
since Flory provides correct number fractions that were derived from
the incorrect equation (27). Note that Ref. \citep{Mizerovskii2013}
agrees with Flory regarding all $w_{\text{X}}$ and does not recognize
this mistake.

\nomenclature{$m_{\text{k}}$}{$k$-th moment of distribution}%

\nomenclature{$E_{\text{X}}$}{normalization constant for computing moments of chain types $X=A,B,m$}%

To simplify notation, let us denote the terms in the square brackets
of equation (\ref{eq:omegam}) to (\ref{eq:omegab}) by $E_{\text{m}}$,
$E_{\text{A}}$, and $E_{\text{B}}$ respectively. We further set
$q=rp^{2}$ and use the moments
\begin{equation}
m_{1}\left(2z+2\right)=\sum_{z=0}^{\infty}\left(2z+2\right)q^{z}=\frac{2}{\left(1-q\right)^{2}}\label{eq:m1}
\end{equation}
\begin{equation}
m_{1}(2z+1)=\sum_{z=0}^{\infty}\left(2z+1\right)q^{z}=\frac{1+q}{\left(1-q\right)^{2}}\label{eq:m2}
\end{equation}
\begin{equation}
m_{2}\left(2z+2\right)=\sum_{z=0}^{\infty}\left(2z+2\right)^{2}q^{z}=4\frac{1+q}{\left(1-q\right)^{3}}\label{eq:m3}
\end{equation}
\begin{equation}
m_{2}(2z+1)=\sum_{z=0}^{\infty}\left(2z+1\right)^{2}q^{z}=\frac{1+6q-q^{2}}{\left(1-q\right)^{3}}.\label{eq:m4}
\end{equation}
The total weight fraction of each termination class is then
\begin{equation}
\text{w}_{\text{m}}=m_{1}\left(2z+2\right)E_{\text{m}}\label{eq:om}
\end{equation}
and for $X=A,B$ there is
\begin{equation}
\text{w}_{\text{X}}=m_{1}(2z+1)E_{\text{X}}\label{eq:oa}
\end{equation}
As a test, we have checked normalization by computing $\text{w}_{\text{m}}+\text{w}_{\text{A}}+\text{w}_{\text{B}}$,
which is indeed unity for our set of equations but not for the equations
provided in Refs. \citep{Flory1936,Mizerovskii2013}. Therefore, dependent
results like weight average DP or polydispersity in these works need
to be questioned.

The weight average DP is computed as $m_{2}/m_{1}$ where the $E_{\text{X}}$
terms for all $X=A,$ $B$, and $m$ cancel out. This yields
\begin{equation}
N_{\text{w,m}}=\frac{m_{2}\left(2z+2\right)}{m_{1}\left(2z+2\right)}=2\left(\frac{1+q}{1-q}\right)\label{eq:Nwm}
\end{equation}
\begin{equation}
N_{\text{w,A}}=N_{\text{w,B}}=\frac{m_{2}\left(2z+1\right)}{m_{1}\left(2z+1\right)}=\frac{1+6q+q^{2}}{1-q^{2}}\label{eq:NwA}
\end{equation}
The last two equations allow to compute the total weight average DP
through
\begin{equation}
N_{\text{w}}=N_{\text{w,m}}\text{w}_{\text{m}}+N_{\text{w,A}}\left(1-\text{w}_{\text{m}}\right).\label{eq:Nw}
\end{equation}
Here the $E_{\text{X}}$ terms do not cancel, and a rather complex
result is obtained that we do not reproduce here. The number average
DP of the mixed terminated chains is given by
\begin{equation}
N_{\text{n,m}}=\frac{2\sum_{z=0}^{\infty}\left(z+1\right)q^{z}}{\sum_{z=0}^{\infty}q^{z}}=\frac{2}{1-q}\label{eq:Nnm}
\end{equation}
while the number average DP of the $A$ and $B$ terminated chains
is less by one,
\begin{equation}
N_{\text{n,A}}=N_{\text{n,m}}-1=\frac{1+q}{1-q}=N_{\text{n,B}},\label{eq:NnA}
\end{equation}
since the weight distribution starts from a single strand and not
from a pair. Finally, the polydispersities are
\begin{equation}
\frac{N_{\text{w,m}}}{N_{\text{n,m}}}=1+q\label{eq:polym}
\end{equation}
\begin{equation}
\frac{N_{\text{w,A}}}{N_{\text{n,A}}}=\frac{N_{\text{w,B}}}{N_{\text{n,B}}}=\frac{1+6q+q^{2}}{\left(1+q\right)^{2}}.\label{eq:polyA}
\end{equation}

\nomenclature{$N_{\text{n,X}}$}{number average DP of $X=A,B,m$ terminated chains}%

\nomenclature{$N_{\text{w,X}}$}{weight average DP of $X=A,B,m$ terminated chains}%

\nomenclature{$q$}{probability for adding one additional pair of strands}%

Note that the first of these polydispersities is equivalent to the
polydispersity of a most probable distribution for $r=1$. The second
result is identical to the polydispersity of an alternating co-polymerization
at full conversion \citep{Lang2019c}, which supports that also the
averages for the $A$ and $B$ terminated chains and all intermediate
steps are correct. Note that Flory \citep{Flory1936} did not compute
weight averages and polydispersity, while Mizerovskii and Padokhin
\citep{Mizerovskii2013} arrive at several expressions that contain
non-integer powers of probabilities like $r$. Such results are obviously
not correct, since the probability $r$ is associated with the existence
of $A$ strands: these either exist or not, but they cannot adopt
any state in between.

\subsection{\label{subsec:Weight-distributions-and-1}Case 2b without rings}

We consider the polymerization of linear strands with two different
reactive groups $A$ and $B$ on either end forming exclusively $AA$
and $BB$ bonds with a probability $p_{\text{A}}$ and $p_{\text{B}}$,
respectively. To compute these conversions, we assume the independence
of the reactions of $A$ and $B$ groups. In the absence of intra-molecular
reactions, these establish separate equilibrium concentrations of
closed stickers according to \citep{Stukhalin2013,Lang2021a}
\begin{equation}
p_{\text{X}}=1-\frac{\left(1+8K_{\text{X}}c_{\text{X}}\right)^{1/2}-1}{4K_{\text{X}}c_{\text{X}}}\label{eq:px}
\end{equation}
where $K_{\text{X}}$ is the reaction constant for $X=A,B$ groups
respectively, and $c_{\text{X}}=c_{\text{A}}=c_{\text{B}}$. Note
that here a factor of $2K$ appears instead of $K$ as in case 1,
since the concentration of $B$ groups does not play any role for
reactions of $A$ groups and vice versa, see also the discussion around
equation (2) of Ref. \citep{Lang2021a}. Total conversion is given
by
\begin{equation}
p=\left(p_{\text{A}}+p_{\text{B}}\right)/2.\label{eq:totalp}
\end{equation}

The derivation below follows closely the steps in the preceding section
concerning case 3. Therefore, we skip here most of the discussion,
except for deviations from case 3.

When picking randomly a chain end, a fraction of 
\begin{equation}
P_{\text{end}}\left(A\right)=\left(1-p_{\text{A}}\right)/\left(2-p_{\text{A}}-p_{\text{B}}\right)\label{eq:P_end}
\end{equation}
of these ends is of type $A$, while a fraction of $1-P_{\text{end}}\left(A\right)$
is of type $B$. The average degree of polymerization of the linear
chains, $N_{\text{n}}$, is the concentration of reactive groups divided
by the concentration of chain ends
\begin{equation}
N_{\text{n}}=\frac{2c_{\text{A}}}{c_{\text{A}}\left(2-p_{\text{A}}-p_{\text{B}}\right)}=\frac{1}{1-p}.\label{eq:Nn-2}
\end{equation}

In analogy to the derivation in the preceding section \emph{Case 3
without rings}, we set $q=p_{\text{A}}p_{\text{B}}$ with $\sum_{z=0}^{\infty}q^{z}=\left(1-q\right)^{-1}$.
$z$ counts again the number of additional pairs of strands beyond
the shortest possible chain of a particular group. In contrast to
the preceding section, the $A$- and $B$-terminated chains consist
now of an even number of strands, while the $m$ terminated chains
contain an odd number of strands. Thus, the arguments previously used
for the mixed terminated chains provide
\begin{equation}
n_{\text{A}}=P_{\text{end}}\left(A\right)p_{\text{B}}\left(1-p_{\text{A}}\right)\sum_{z=0}^{\infty}q^{z}\label{eq:nA}
\end{equation}
\[
=\frac{p_{\text{B}}\left(1-p_{\text{A}}\right)^{2}}{\left(2-p_{\text{A}}-p_{\text{B}}\right)\left(1-p_{\text{A}}p_{\text{B}}\right)},
\]
\begin{equation}
n_{\text{B}}=\frac{p_{\text{A}}\left(1-p_{\text{B}}\right)^{2}}{\left(2-p_{\text{A}}-p_{\text{B}}\right)\left(1-p_{\text{A}}p_{\text{B}}\right)},\label{eq:nB-1}
\end{equation}
while, conversely, we find
\begin{equation}
n_{\text{m}}=\frac{2\left(1-p_{\text{A}}\right)\left(1-p_{\text{B}}\right)}{\left(2-p_{\text{A}}-p_{\text{B}}\right)\left(1-p_{\text{A}}p_{\text{B}}\right)}.\label{eq:nm-1}
\end{equation}
We obtain for the number fraction distributions
\begin{equation}
n_{\text{A,z}}=p_{\text{A}}^{z-1}p_{\text{B}}^{z}\frac{\left(1-p_{\text{A}}\right)^{2}}{\left(2-p_{\text{A}}-p_{\text{B}}\right)},\label{eq:naz}
\end{equation}
\begin{equation}
n_{\text{B,z}}=p_{\text{B}}^{z-1}p_{\text{A}}^{z}\frac{\left(1-p_{\text{B}}\right)^{2}}{\left(2-p_{\text{A}}-p_{\text{B}}\right)},\label{eq:nbz}
\end{equation}
and
\begin{equation}
n_{\text{m,z}}=\left(p_{\text{A}}p_{\text{B}}\right)^{z-1}\frac{2\left(1-p_{\text{A}}\right)\left(1-p_{\text{B}}\right)}{\left(2-p_{\text{A}}-p_{\text{B}}\right)}.\label{eq:nmz-1}
\end{equation}
Let us again use $q=p_{\text{A}}p_{\text{B}}$ to simplify the notation
for the higher moments of the distribution. With an adapted version
of equation (\ref{eq:omega}) where even and odd terms are mutually
exchanged for $m\rightleftharpoons A,B$, we obtain for the weight
fraction distributions that these correspond to 
\begin{equation}
\text{w}_{\text{A,z}}=\left(2z+2\right)q^{z-1}\left[p_{\text{B}}\left(1-p_{\text{A}}\right)^{2}/2\right],\label{eq:oaz}
\end{equation}
\begin{equation}
\text{w}_{\text{B,z}}=\left(2z+2\right)q^{z-1}\left[p_{\text{A}}\left(1-p_{\text{B}}\right)^{2}/2\right],\label{eq:obz}
\end{equation}
and
\begin{equation}
\text{w}_{\text{m,z}}=\left(2z+1\right)q^{z-1}\left[\left(1-p_{\text{A}}\right)\left(1-p_{\text{B}}\right)\right].\label{eq:omz}
\end{equation}
The terms in the square brackets are denoted below by $E_{\text{X}}$
with $X=A,B,m$ accordingly. These do not change for higher order
averages. We further use the moments defined in equation (\ref{eq:m1})
to (\ref{eq:m4}). The weight fractions of chains in each termination
class are then
\begin{equation}
\text{\ensuremath{\text{w}_{\text{A}}=}}m_{1}\left(2z+2\right)E_{\text{A}},\label{eq:oA}
\end{equation}
\begin{equation}
\text{\ensuremath{\text{w}_{\text{B}}=}}m_{1}\left(2z+2\right)E_{\text{B}},\label{eq:oB}
\end{equation}
and
\begin{equation}
\text{\ensuremath{\text{w}_{\text{m}}=}}m_{1}\left(2z+1\right)E_{\text{m}}.\label{eq:om-1}
\end{equation}
As above, the weight average degrees of polymerization are obtained
by the ratio of the moments $m_{\text{2}}/m_{1}$ of the corresponding
even and odd terms. This yields
\begin{equation}
N_{\text{w},\text{A}}=2\frac{1+q}{1-q}=N_{\text{w,B}}\label{eq:Nwa}
\end{equation}
and
\begin{equation}
N_{\text{w,m}}=\frac{1+6q+q^{2}}{1-q^{2}}\label{eq:Nwm-1}
\end{equation}
with a weight average DP in the full sample of
\begin{equation}
N_{\text{w}}=N_{\text{w,m}}\text{w}_{\text{m}}+N_{\text{w,A}}\left(1-\text{w}_{\text{m}}\right).\label{eq:Nw-1}
\end{equation}
The number average degrees of polymerization are
\begin{equation}
N_{\text{n,A}}=\frac{2\sum_{z=0}^{\infty}\left(z+1\right)q^{z}}{\sum_{z=0}^{\infty}q^{z}}=\frac{2}{1-q}=N_{\text{n,B}}\label{eq:NnA-1}
\end{equation}
\begin{equation}
N_{\text{n,m}}=N_{\text{n,A}}-1=\frac{1+q}{1-q}\label{eq:Nnm-1}
\end{equation}
and polydispersities are
\begin{equation}
\frac{N_{\text{w,A}}}{N_{\text{n,A}}}=\frac{N_{\text{w,B}}}{N_{\text{n,B}}}=1+q\label{eq:Nwa-1}
\end{equation}
\begin{equation}
\frac{N_{\text{w,m}}}{N_{\text{n,m}}}=\frac{1+6q+q^{2}}{\left(1+q\right)^{2}}.\label{eq:Nwm-2}
\end{equation}
Altogether, the higher moments of the distributions are equivalent
to case 3. However, the $A$- and $B$- terminated chains contain
now an even number of strands and have the corresponding higher order
averages, while the $m$-terminated chains contain an odd number of
strands with the corresponding higher order averages.

\printnomenclature{}

\bibliographystyle{achemso}
\bibliography{library}

\begin{figure*}
\includegraphics[width=1\textwidth]{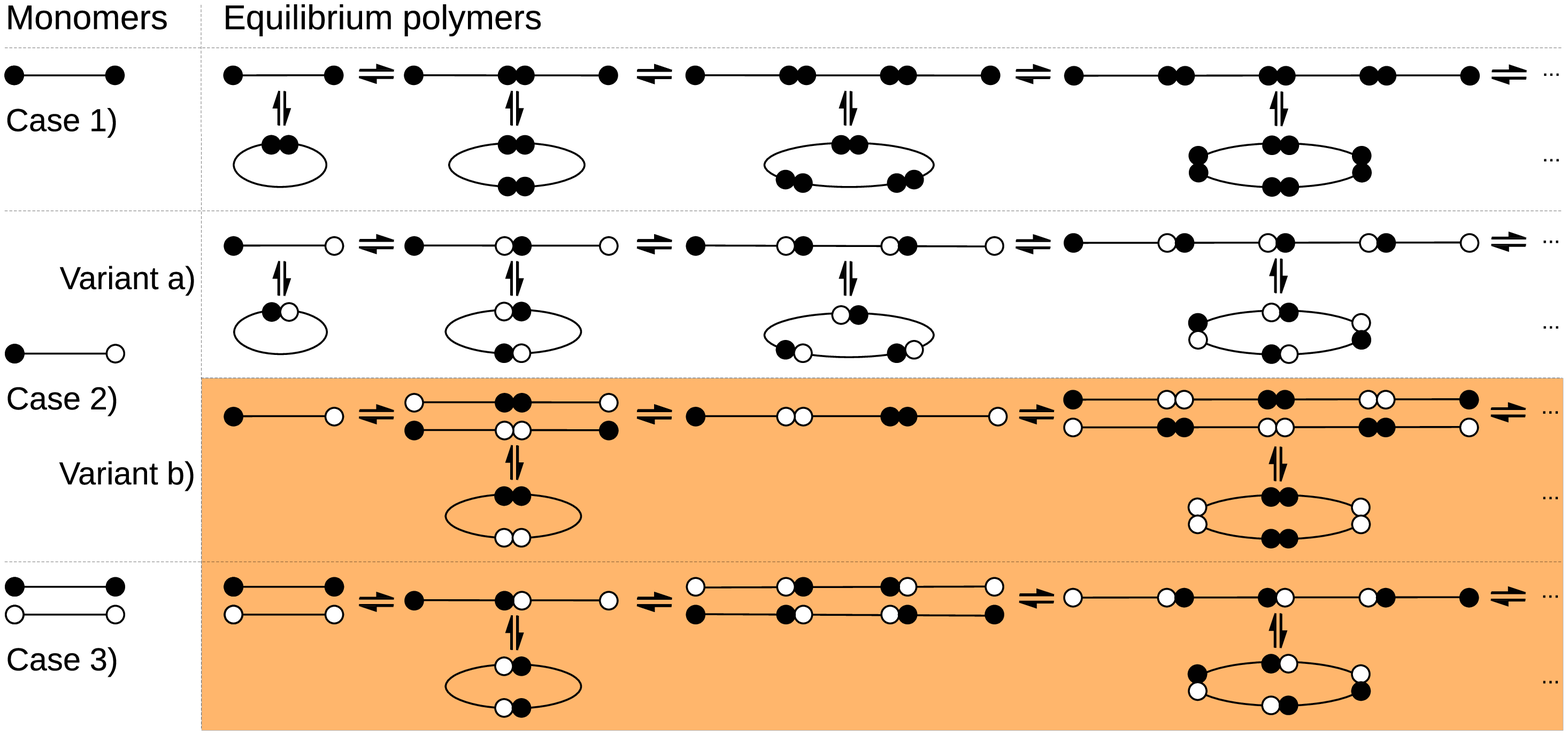}
\end{figure*}

\end{document}